\documentclass[useAMS,usenatbib,usegraphicx]{mn2e}
\usepackage{amsmath}
\usepackage{aas_macros}
\usepackage[british]{babel}
\usepackage{multirow}
\usepackage{url}
\bibpunct{(}{)}{,}{a}{}{,} 
\newcommand{\lt}{<}

\newcommand{\pms}{\texttt{pre-MS}}

\newcommand{\eos}{\texttt{EOS}}
\newcommand{\ml}{$\alpha_\mathrm{ML}$}
\newcommand{\prosecco}{\textsc{PROSECCO}}
\newcommand{\ldb}{\texttt{LDB}}
\newcommand{\bc}{\texttt{BC}}
\newcommand{\bcs}{\texttt{BCs}}
\newcommand{\libur}{$^7$Li-burning}
\title[Cumulative uncertainties in \ldb{} age]{Cumulative theoretical uncertainties in lithium depletion boundary age}
\author[E. Tognelli, P. G. Prada Moroni, \& S. Degl'Innocenti]
{E. Tognelli$^{1,2}$\thanks{e-mail: tognelli$@$df.unipi.it}, P.G. Prada Moroni$^{2,3}$\thanks{e-mail: prada$@$df.unipi.it}, 
S. Degl'Innocenti$^{2,3}$\\
$^{1}$Department of Physics, University of Roma Tor Vergata, Via della Ricerca Scientifica 1, 00133, Roma, Italy\\
$^{2}$INFN, Section of Pisa, Largo Bruno Pontecorvo 3, 56127, Pisa, Italy\\
$^{3}$Department of Physics `E.Fermi', University of Pisa, Largo Bruno Pontecorvo 3, 56127, Pisa, Italy
}
\begin{document}
\date{Accepted 2015 March 13. Received 2015 March 13; in original form 2014 July 25}
\pagerange{\pageref{firstpage}--\pageref{lastpage}} \pubyear{2015}
\maketitle
\label{firstpage}
%
\begin{abstract}
We performed a detailed analysis of the main theoretical uncertainties affecting the age at the lithium depletion boundary (\ldb). To do that we computed almost 12\,000 pre-main sequence models with mass in the range [0.06, 0.4] M$_\odot$ by varying input physics (nuclear reaction cross-sections, plasma electron screening, outer boundary conditions, equation of state, and radiative opacity), initial chemical elements abundances (total metallicity, helium and deuterium abundances, and heavy elements mixture), and convection efficiency (mixing length parameter, \ml). As a first step, we studied the effect of varying these quantities individually within their extreme values. Then, we analysed the impact of simultaneously perturbing the main input/parameters without an a priori assumption of independence. Such an approach allowed us to build for the first time the cumulative error stripe, which defines the edges of the maximum uncertainty region in the theoretical \ldb{} age.

We found that the cumulative error stripe is asymmetric and dependent on the adopted mixing length value. For \ml{} = 1.00, the positive relative age error ranges from 5 to $15\%$, while for solar-calibrated mixing length, the uncertainty reduces to $5- 10\%$. A large fraction of such an error ($\approx 40\%$) is due to the uncertainty in the adopted initial chemical elements abundances.
\end{abstract}
\maketitle
%
\section{Introduction}
Lithium burning via the proton capture $^7$Li(p, $\alpha$)$\alpha$ becomes efficient in stellar conditions as temperature rises above $2.5 \times 10^6$ K. In the mass range 0.06-0.4 M$_\odot$ (the exact values depending on the chemical composition), such a temperature is already reached during the pre-main sequence (\pms) phase allowing lithium to be completely destroyed in fully convective structures. Since the larger the mass and the higher the rate of central temperature increase and, consequently, the earlier the onset of lithium burning, the age at which lithium is fully depleted strongly depends on mass. Thus, in stellar associations or clusters with ages between 15 and 350 Myr, one would expect to observe a sharp transition in the very low mass regime between stars with and without photospheric lithium. Such a transition, usually called  lithium depletion boundary (\ldb), is an age indicator \citep[see e.g.][]{dantona94}. 

The age at the \ldb{} can be provided only by theoretical stellar evolutionary models. The accuracy of their predictions relies on the adopted input physics (e.g. equation of state, radiative opacity, nuclear reaction cross-sections, etc.), chemical composition (initial metallicity, helium and deuterium abundance, heavy element mixture, etc.), and the numerical scheme describing macroscopic physical processes (convection, rotation, etc\dots). All these ingredients are affected by not negligible uncertainties, which translate into a theoretical error in the \ldb{} age estimate.

Given the primary importance of evaluating the age of young stellar clusters, several studies have been devoted to quantify the main theoretical and observational uncertainties affecting the \ldb{} method. A first idea of the theoretical uncertainty is provided by comparing models computed by different groups \citep[see e.g.][]{jeffries01b}. However such an approach is neither able to quantify the individual uncertainty sources nor able to fully exploit the whole theoretical uncertainty, as several input physics are in common among the different evolutionary codes.

A better and more detailed way to proceed is to vary a single ingredient at a time and to study its impact on the \ldb{} age \citep[see e.g.][]{bildsten97,ushomirsky98,burke04}. The weakness of this approach is that it is unable to quantify the interactions between the different uncertainty sources. 

A more robust procedure, although much more computationally expensive, consists in varying simultaneously the input physics, parameters, and chemical abundances adopted in stellar computations within their current uncertainty range. Such an estimate of the theoretical uncertainty affecting the \ldb{} ages is still lacking and, thus, we decided to provide it. We addressed also for the first time the analysis of the impact of the current uncertainty in the adopted chemical composition on the \ldb{} age estimates. To do that, we followed the same procedure adopted in our previous studies on the cumulative physical uncertainty affecting low-mass star models from the main sequence to the He-burning phase \citep{valle13a,valle13b}, that is, a systematic and simultaneous variation of the main input physics on a fixed grid. 

The number of models required to fully cover the whole parameter space is huge. For this reason, as a first step, we started to study the effect of individual input perturbations to discriminate between negligible and not negligible error sources. In Sect. \ref{sec:var_indi_fis} we computed sets of perturbed models by changing one input physics at a time keeping fixed the chemical composition and in Sect. \ref{sec:var_indi_chm} we computed sets of perturbed models by varying individually the initial metallicity, helium and deuterium abundances, and heavy elements mixture at a time at fixed input physics. 
Finally, in Sect. \ref{sec:global} we computed sets of models by simultaneously varying all the ingredients that have a not negligible effect on the \ldb{} age. Section \ref{sec:conclusions} reports the main conclusions. This work required the computation of about $12\,000$ \pms{} tracks. 

\section{The reference set of models}
\label{sec:reference}
We computed the models with the most recent release of the \prosecco{} (\textbf{P}isa \textbf{R}aphson-Newt\textbf{O}n \textbf{S}tellar \textbf{E}volution \textbf{C}omputation \textbf{CO}de) stellar evolution code derived from the \textsc{FRANEC} code \citep[see e.g.][]{deglinnocenti08}. A detailed description of the standard models can be found in \citet{tognelli11}, \citet{tognelli12} and  \citet{dellomodarme12}; here we limit the discussion only to the variations with respect to the previous version.

The reference models are standard \pms{} tracks, which do not take into account accretion, rotation, and magnetic fields. The evolution starts from a fully convective and homogeneous model on the Hayashi track, with a central temperature $T_\mathrm{c} \approx 10^5$ K, low enough for deuterium burning to be completely inefficient. Although the initial model is not physically realistic, as it does not result from a self consistent protostar evolution, we checked that varying the initial central temperature (hence radius) in the range [$10^5$, $10^6$] K does not affect the \ldb.

We adopted the 2006 release of the \textsc{OPAL} equation of state \citep[\textsc{OPAL06} \eos; ][]{rogers02}. However, in order to compute objects less massive than 0.1 M$_\odot$, in the current version of the code we included also the \eos{} by \citet[][\textsc{SCVH95}]{saumon95}, for temperatures and densities not covered by the \textsc{OPAL06} \eos. 

We used the same radiative \citep[\textsc{OPAL 2005} and ][]{ferguson05} and conductive \citep{potekhin99,shternin06} opacities as in \citet{tognelli11} but for a different solar heavy elements mixture, namely the \citet[][\texttt{AS09}]{asplund09}. 

Nuclear reaction rates relevant for the \pms{} evolution have been taken from the \texttt{NACRE} compilation \citep{nacre}, with the exception of p(p,e$^+\nu$)$^2$H \citep{marcucci13,tognelli15}, $^2$H(p,$\gamma$)$^3$He \citep{descouvemont04}, $^2$H($^2$H,p)$^3$H and $^2$H($^2$H,n)$^3$He \citep{tumino14},$^6$Li(p,$^3$He)$\alpha$ \citep{lamia13}, and $^7$Li(p,$\alpha$)$\alpha$ \citep{lamia12}. Nuclear reaction rates between bare nuclei have been corrected to account for plasma electron screening. We implemented the weak \citep{salpeter54}, weak$-$intermediate$-$strong \citep{dewitt73,graboske73}, and strong \citep{itoh77,itoh79} screening. Then, an adaptive procedure determines the most suitable screening factor to be used in each mass-shell of the model. For what concerns the burning of the light elements relevant for this work (mainly $^2$H and $^7$Li), the intermediate/strong screening is generally adopted.

Outer boundary conditions (\bcs) have been obtained by the detailed atmospheric models by \citet[][\texttt{BH05}]{brott05} as discussed in \citet{tognelli11}. Convection is treated according to the mixing length theory \citep{bohm58} following the formalism described in \citet{cox}. Usually the solar-calibrated mixing length value (\ml) is adopted for different stellar masses and evolutionary phases. However, there are no compelling theoretical arguments for such a choice \citep[see e.g.,][]{ludwig99,trampedach07}. In particular, there are several hints that in \pms{} phase the superadiabatic convection might be less efficient \citep[see e.g. ][and references therein]{dantona03,stassun04,mathieu07,stassun08,somers14}. Using our models a good agreement with eclipsing binaries and surface lithium abundances in young clusters is obtained with \ml$\sim 1$ \citep{gennaro12, tognelli12}. We adopted \ml = 1.00 in our reference models. For the sake of completeness, in the following we will show also the effect of using our solar-calibrated value \ml = 1.74\footnote{The solar model has been computed using an iterative procedure to adjust the initial helium abundance, metallicity, and mixing length parameter in our 1 M$_\odot$ stellar model, in order to reproduce, within a given numerical tolerance ($\lt 10^{-4}$), at the age of the Sun (4.57 Gyr) its observed radius, luminosity, and $(Z/X)_\mathrm{ph}$.}.

The reference set of models have been computed adopting the solar heavy element mixture by \citet{asplund09} and a helium-to-metal enrichment ratio $\Delta Y/\Delta Z = 2$ \citep{casagrande07}. With this choices, [Fe/H] = $+0.0$ translates into an initial helium abundance $Y=0.274$ and a global metallicity $Z=0.013$ (see Sect. \ref{sec:var_indi_chm}). The initial abundances of the light elements are the same used in \citet{tognelli12}.

\subsection{Lithium depletion boundary}
We computed 36 \pms{} tracks in the mass range [0.06, 0.40] M$_\odot$, in which lithium is completely destroyed in a fully convective object. We focused on this mass range because for the chosen value [Fe/H] = $+0.0$ objects less massive than about 0.06 M$_\odot$ never reach the temperature required to destroy lithium, while for M $\ga 0.4$ M$_\odot$ a radiative core develops before lithium is completely depleted. 

We defined the \ldb{} as the model at which lithium abundance in mass is reduced by a factor of 100 with respect to the original one. In order to compute a reliable lithium depletion evolution and, consequently \ldb{} ages, the time-step adopted in model computations must be chosen with care. A time-step simply tuned to obtain an accurate \pms{} evolution would result in a too crude time resolution and in an underestimate of the lithium depletion at a given age, which in turn would lead to an overestimate of the \ldb{} age \citep{piau02,burke04}. This is the consequence of \libur{} occurring over temporal-scales much shorter than the \pms{} evolutionary ones. Being aware of this, in the routine that provides the time-step for each iteration we introduced an additional condition that constraints the time-step to be short enough to lead to a relative lithium abundance variation lower than $0.1\%$. Moreover, such a choice allows us to obtain also a good numerical resolution in the $^7$Li abundance to precisely identify the \ldb{} point.

\begin{figure}
	\centering
	\includegraphics[width=\columnwidth]{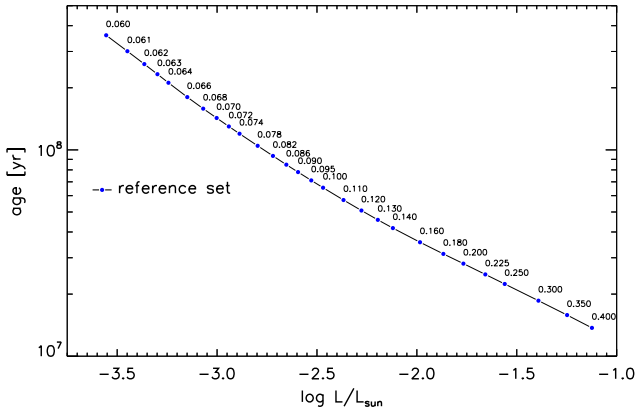}
	\caption{Age at which lithium is depleted by a factor of 100 (\ldb) as a function of luminosity for our reference set of models with $Z=0.0130$, $Y=0.274$, $X_\mathrm{d}=2\times10^{-5}$, and \ml = 1.00. Stellar masses (in M$_\odot$) are overplotted.}
	\label{fig:ref}
\end{figure}
Fig. \ref{fig:ref} shows our reference \ldb{} curve in the ($\log L/$L$_\odot$, age) plane obtained from the reference set of models. The complete set is listed in Table \ref{tab:refe} with the corresponding age, $\log L/$L$_\odot$, $\log T_\mathrm{eff}$, radius, and surface gravity values at the \ldb{} for each mass\footnote{Table \ref{tab:refe} along with \ldb{} ages for several [Fe/H] values is available in electronic form at the url:\\ \url{http://astro.df.unipi.it/stellar-models/ldb}}.

\begin{table}
	\centering
	\caption{\ldb{} quantities for the reference set of models with $Z~=~0.0130$, $Y=0.274$, $X_\mathrm{d}=2\times10^{-5}$, and \ml = 1.00.}
	\label{tab:refe}
	\begin{tabular}{lccccc}
	\hline
	$M_\mathrm{tot}$ & Age & $\log L/\mathrm{L}_\odot$ & $\log R/\mathrm{R}_\odot$ & $\log g$ & $\log T_\mathrm{eff}$ \\
	(M$_\odot$) & (Myr) & & & (cm s$^{-2})$ & (K)\\
	\hline
	\hline	
	$0.060$ & $359$ & $ -3.555$ & $ -0.972$ & $  5.160$ & $  3.359$ \\
	$0.061$ & $300$ & $ -3.449$ & $ -0.954$ & $  5.132$ & $  3.377$ \\
	$0.062$ & $260$ & $ -3.364$ & $ -0.939$ & $  5.108$ & $  3.390$ \\
	$0.063$ & $232$ & $ -3.298$ & $ -0.925$ & $  5.088$ & $  3.400$ \\
	$0.064$ & $211$ & $ -3.243$ & $ -0.913$ & $  5.070$ & $  3.407$ \\
	$0.065$ & $194$ & $ -3.195$ & $ -0.902$ & $  5.055$ & $  3.414$ \\
	$0.066$ & $180$ & $ -3.149$ & $ -0.891$ & $  5.040$ & $  3.420$ \\
	$0.068$ & $158$ & $ -3.069$ & $ -0.872$ & $  5.015$ & $  3.430$ \\
	$0.070$ & $142$ & $ -3.002$ & $ -0.855$ & $  4.993$ & $  3.439$ \\
	$0.072$ & $129$ & $ -2.941$ & $ -0.839$ & $  4.972$ & $  3.446$ \\
	$0.074$ & $119$ & $ -2.888$ & $ -0.824$ & $  4.954$ & $  3.452$ \\
	$0.075$ & $115$ & $ -2.864$ & $ -0.817$ & $  4.946$ & $  3.454$ \\
	$0.076$ & $111$ & $ -2.840$ & $ -0.809$ & $  4.937$ & $  3.456$ \\
	$0.077$ & $107$ & $ -2.817$ & $ -0.803$ & $  4.929$ & $  3.459$ \\
	$0.078$ & $104$ & $ -2.797$ & $ -0.796$ & $  4.922$ & $  3.461$ \\
	$0.079$ & $101$ & $ -2.776$ & $ -0.790$ & $  4.915$ & $  3.462$ \\
	$0.080$ & $ 98$ & $ -2.757$ & $ -0.783$ & $  4.908$ & $  3.464$ \\
	$0.082$ & $ 93$ & $ -2.720$ & $ -0.771$ & $  4.895$ & $  3.467$ \\
	$0.084$ & $ 88$ & $ -2.686$ & $ -0.760$ & $  4.882$ & $  3.470$ \\
	$0.086$ & $ 84$ & $ -2.653$ & $ -0.749$ & $  4.869$ & $  3.473$ \\
	$0.088$ & $ 81$ & $ -2.623$ & $ -0.738$ & $  4.858$ & $  3.475$ \\
	$0.090$ & $ 78$ & $ -2.595$ & $ -0.728$ & $  4.847$ & $  3.477$ \\
	$0.095$ & $ 71$ & $ -2.529$ & $ -0.703$ & $  4.822$ & $  3.481$ \\
	$0.100$ & $ 65$ & $ -2.469$ & $ -0.681$ & $  4.799$ & $  3.485$ \\
	$0.110$ & $ 57$ & $ -2.367$ & $ -0.640$ & $  4.759$ & $  3.490$ \\
	$0.120$ & $ 50$ & $ -2.277$ & $ -0.603$ & $  4.723$ & $  3.494$ \\
	$0.130$ & $ 45$ & $ -2.195$ & $ -0.570$ & $  4.691$ & $  3.498$ \\
	$0.140$ & $ 41$ & $ -2.120$ & $ -0.539$ & $  4.662$ & $  3.501$ \\
	$0.160$ & $ 35$ & $ -1.984$ & $ -0.484$ & $  4.611$ & $  3.508$ \\
	$0.180$ & $ 31$ & $ -1.867$ & $ -0.437$ & $  4.568$ & $  3.514$ \\
	$0.200$ & $ 28$ & $ -1.767$ & $ -0.397$ & $  4.532$ & $  3.518$ \\
	$0.225$ & $ 24$ & $ -1.657$ & $ -0.352$ & $  4.493$ & $  3.523$ \\
	$0.250$ & $ 22$ & $ -1.560$ & $ -0.312$ & $  4.459$ & $  3.527$ \\
	$0.300$ & $ 18$ & $ -1.392$ & $ -0.243$ & $  4.401$ & $  3.535$ \\
	$0.350$ & $ 15$ & $ -1.248$ & $ -0.185$ & $  4.353$ & $  3.542$ \\
	$0.400$ & $ 13$ & $ -1.124$ & $ -0.136$ & $  4.312$ & $  3.549$ \\
	\hline
	\end{tabular}
\end{table}

We also provide a fit of the $\log \mathrm{age}[\mathrm{yr}]$ - $\log L/$L$_\odot$ curve for the reference set of models using a third-order polynomial fit\footnote{`$\log$' stands for the base-10 logarithm.},
\begin{eqnarray}
\log \mathrm{age}[\mathrm{yr}] &=& a_0 + a_1\times\log L/\rmn{L}_\odot + a_2\times(\log L/\rmn{L}_\odot)^2 \nonumber\\
 && + a_3\times(\log L/\rmn{L}_\odot)^3 \nonumber
\end{eqnarray}
The fit parameters $a_0,\,a_1,\,a_2,$ and $a_3$ are given in Table \ref{tab:fit}; the mean accuracy of the fit is better than $1\%$ over the whole selected luminosity range. 
\begin{table}
	\centering
	\caption{Coefficients of the third-order polynomial fit of the \ldb{} age versus luminosity.}
	\label{tab:fit}
	\begin{tabular}{ccccc}
	\hline
	$a_0$ & $a_1$ & $a_2$ & $a_3$\\
	\hline
	\hline
	$6.6408$ & $-0.46379$ & $-0.031392$ & $-0.014906$ \\
	\hline
	\end{tabular}
\end{table}

\section{Individual input physics uncertainties}
\label{sec:var_indi_fis}
The results of stellar model computations depend on the adopted input physics, such as nuclear reaction cross-sections, plasma electron screening, radiative opacity, equation of state (\eos), and \bcs. All these ingredients are still affected by a not negligible uncertainty which directly translates into an uncertainty in stellar model outcomes \citep{valle13a,valle13b}. In this section, we focus only on the input physics that might affect the lithium burning in \pms{} of very low mass stars. 

As a first estimate of the physical uncertainty impact on the \ldb{} age, we varied a single input physics at a time keeping all the others fixed \citep[see][ for a similar investigation]{burke04}. More in detail, when the error on a given input physics was available, we computed two additional sets of perturbed models by adopting respectively the highest and lowest value of the input physics given by its uncertainty. Then, the \ldb{} ages provided by the perturbed models have been compared with the reference one. Unfortunately, not for all the analysed input physics an uncertainty evaluation is available; this is for example the case of the \eos{} and outer \bcs. Each of these cases has been conveniently treated, as discussed in the following sub-sections, by substituting the adopted tables. Table \ref{tab:err_fis} lists the analysed input physics with the related assumed uncertainty/range of variation, when present, or the alternative input physics. Note that, only in the case of the nuclear reaction cross-sections the listed errors are $1\sigma$, while in the other cases they represent the extreme values of the variability region.

Where not explicitly stated, all the models have been computed for the reference chemical composition and mixing length parameter, as described in Sect. \ref{sec:reference}.

Some of the cases analysed in the following sub-sections have been already discussed in \citet{burke04}. However, a detailed comparison is difficult because they adopted input physics different from ours and they recalibrated the mixing length parameter and initial helium abundance on the Sun for each perturbed set of models. A variation of the initial $Y$ and/or \ml{} in the perturbed models due to the solar re-calibration might partially counterbalance or increase the effect induced by the sole variation of the analysed quantity. For this reason, we preferred to show the contribution on the \ldb{} age of the sole perturbed quantity with all the other parameters fixed. Moreover, the use of a solar calibrated \ml{} does not guarantee a better agreement with radii of low \pms{} stars, where the lithium is actually depleted. Indeed, several papers have shown that in these stars the superadiabatic convection is much less efficient than in the Sun and that a proper value of \ml{} is of the order of 1 \citep[see e.g.][ and references therein]{ventura98,dantona03,landin06,gennaro12,tognelli12}.

\begin{table}
\centering
\caption{Input physics varied in the computation of perturbed stellar models and their assumed uncertainty or range of variation (see text). The flag `yes' specifies the quantities taken into account in the cumulative uncertainty calculation (see Sect. \ref{sec:global}).}
\label{tab:err_fis}
\begin{tabular}{lcc}
\hline
Quantity & Error & Global\\
\hline
\hline
$^2$H(p,$\gamma$)$^3$He reaction rate & $\pm 3\%$ & No\\
$^2$H($^2$H,n)$^3$He reaction rate & $\pm 5\%$ & No\\
$^2$H($^2$H,p)$^3$H reaction rate& $\pm 5\%$ & No\\
$^7$Li(p,$\alpha$)$\alpha$ reaction rate& $\pm 10\%$ & Yes\\
Electron screening(p+$^7$Li) & $+50\%,\,+100\%$ & No\\
$\overline{\kappa}_\mathrm{rad}$ & $\pm 5\%$ & No\\
$\tau_\mathrm{ph}$ & $2/3,\, 100$ & Yes \\
\bcs\,$^{(a)}$ & \texttt{AHF11}, \texttt{KS66} & No\\
\eos\,$^{(b)}$ & \textsc{OPAL06}, \textsc{FreeEOS08}, & No\\
& \textsc{SCVH95}\\
\hline
\end{tabular}

\medskip
\flushleft
$^{(a)}$ \texttt{AHF11}: \citet{allard11}; \texttt{KS66}: \citet{krishna66}\\
$^{(b)}$ \textsc{OPAL06}: \citet{rogers02}; \textsc{FreeEOS08}: \citet{irwin08}; \textsc{SCVH95}: \citet{saumon95}\\
The \bcs/\eos{} have been varied by using tables provided by different authors, because a proper evaluation of the uncertainty is lacking.
\end{table}

\subsection{Nuclear cross-sections}
Since we are interested in \ldb, the only nuclear reactions that might have an effect are those that take place before or during the \libur. Such reactions are: $^2$H(p,$\gamma$)$^3$He, $^2$H($^2$H,p)$^3$H,  $^2$H($^2$H,n)$^3$He, and $^7$Li(p,$\alpha$)$\alpha$. 

We adopted the following uncertainties in the quoted reactions: $\pm3\%$ for the p+$^2$H reaction \citep[][for temperatures of about $10^6$ K, typical of d-burning]{descouvemont04}, $\pm5\%$ for the $^2$H+$^2$H channels \citep{tumino14}, and $\pm10\%$ for the $^7$Li+p reaction \citep{lamia12}.

Notwithstanding the crucial role played by deuterium burning during the early \pms{} evolution, due to the slight variation of the reaction rates within the current uncertainties, the effect on the \ldb{} age is completely negligible. For this reason we do not show the related plots.
\begin{figure}
	\centering
	\includegraphics[width=\columnwidth]{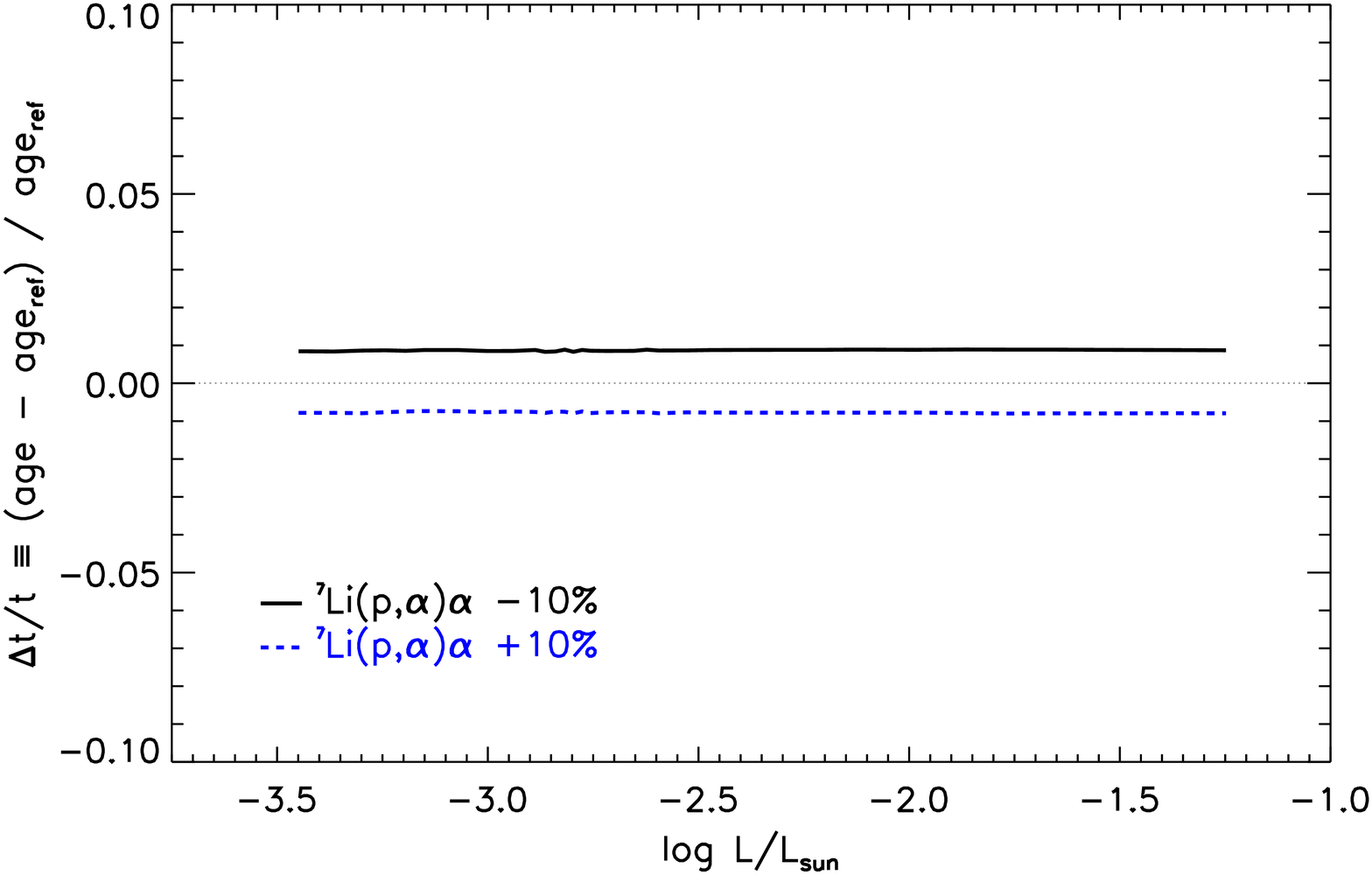}
	\caption{Relative age difference at \ldb{} as a function of luminosity between the reference set of models and the sets with perturbed $^7$Li(p,$\alpha$)$\alpha$ cross-section.}
	\label{fig:lip}
\end{figure}
Fig. \ref{fig:lip} shows the relative age difference, i.e. (perturbed model $-$ reference model)/reference model versus luminosity, due to the uncertainty in the $^7$Li(p,$\alpha$)$\alpha$ cross-section. The relative age difference, about $\pm1\%$, is independent of the mass over the whole selected mass range. An increase of the cross section results in a higher $^7$Li-burning efficiency at a given temperature, which leads to a more rapid $^7$Li depletion and to a lower \ldb{} age at a fixed luminosity. This effect, analysed here for the first time, is small but systematic. We computed also reference and perturbed models adopting the solar-calibrated mixing length parameter (i.e. \ml{}=1.74), verifying that the relative \ldb{} age differences are completely unaffected by a variation of \ml.

\subsection{Plasma electron screening}
\label{sec:sk}
An important point to discuss is the effect of the electron screening on \libur. Plasma electrons around the  interacting nuclei reduce the effective Coulombian repulsion enhancing the reaction rate by a factor $f_{\mathrm{pl}}$ \citep[see e.g.][]{salpeter54,dewitt73,graboske73}. A similar effect, due to atomic electrons, is present in the measurements performed in the laboratory. To this regard, there are hints that  atomic electron screening measured in laboratory is systematically lower ($\sim 1/2$) than the theoretical expectations \citep[see e.g.][and references therein]{pizzone10}.

It is not yet clear whether theoretical computations of plasma electron screening are affected by a similar problem \citep[see e.g.][]{castellani96}. Moreover, at the moment a reliable estimate of such an uncertainty source is lacking. However, the effect of modifying the plasma screening directly affects the reaction rate. For what concerns the \ldb{} age, the uncertainty in plasma screening introduces a systematic effect on the central temperature value at which a reaction becomes efficient, thus reflecting eventually on the \ldb{} age. Given such a situation, it might be useful to check the effect of a variation of $f_{\mathrm{pl}}$ in the $^7$Li(p,$\alpha$)$\alpha$ reaction. Given the lack of a solid estimate of the uncertainty in $f_\mathrm{pl}$(Li), the best thing we can do is assuming a maximum discrepancy between the predicted and real electron plasma screening efficiency similar to that observed in the laboratory; thus, we computed two sets of perturbed models with $f_{\mathrm{pl}}$(Li) increased by a factor 1.5 and 2.0.

\begin{figure}
	\centering
	\includegraphics[width=\columnwidth]{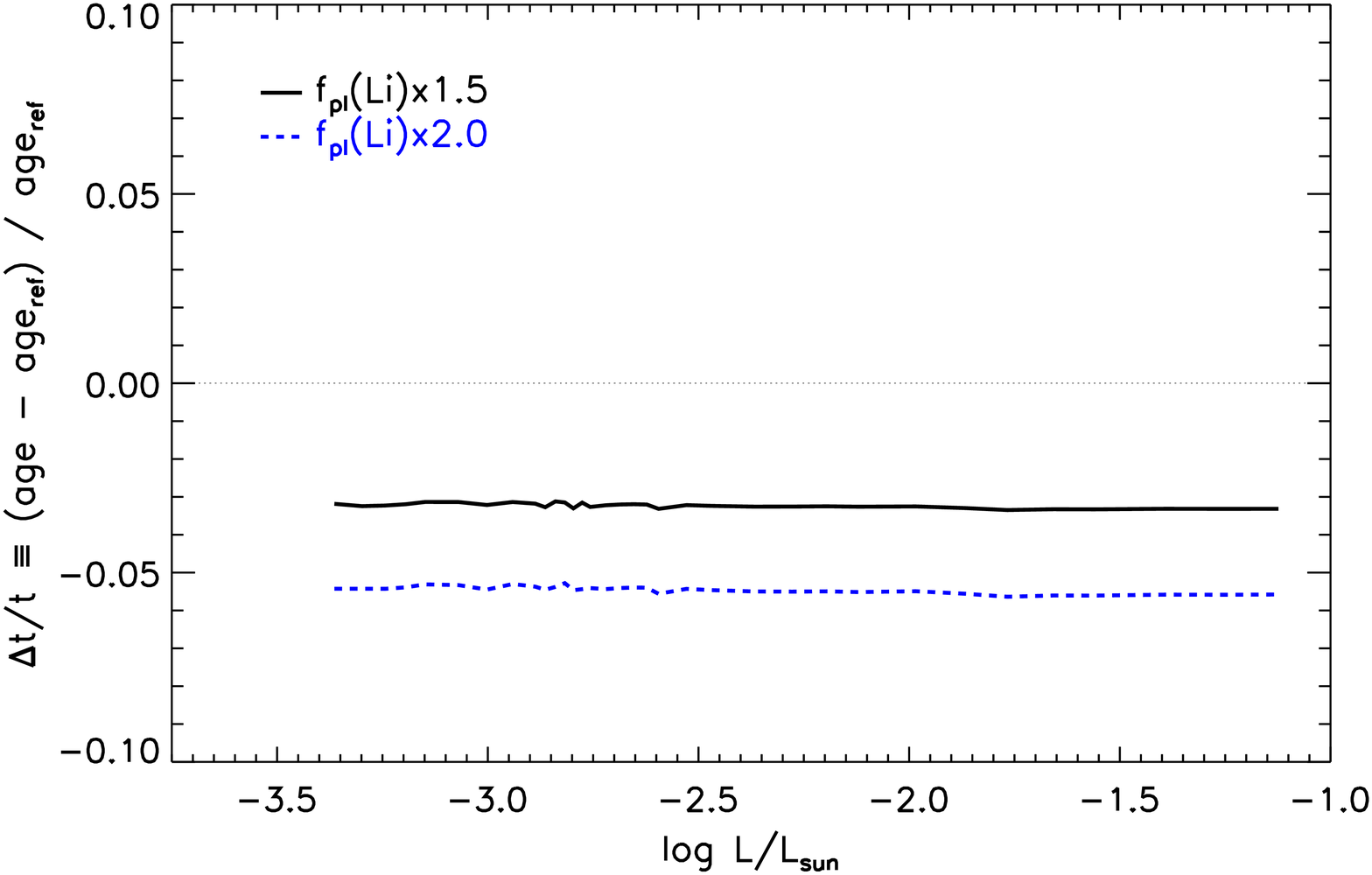}
	\caption{Relative age difference at \ldb{} as a function of luminosity between the reference set of models and the sets with the labelled variations of the electron screening factor.}
	\label{fig:sk}
\end{figure}
Fig. \ref{fig:sk} shows the relative age difference versus luminosity due to the variation of the $^7$Li(p,$\alpha$)$\alpha$ plasma electron screening. Increasing the electron screening leads to higher reaction rates and hence to lower \ldb{} ages at a given luminosity. An enhancement by a factor of $1.5$ and $2.0$ produces, respectively, a relative age decrease of the order of $3\%$ and $5.5\%$ at a fixed luminosity. These results do not change if \ml{}=1.74 is adopted in place of \ml{}=1.00 to compute both the reference and perturbed sets of models.

The plasma electron screening variation, although never analysed before, has a significant effect on the \ldb{} age. However, there are not yet robust estimates of the uncertainty in the $^7$Li(p,$\alpha$)$\alpha$ plasma screening factor, so present computations are intended to only give an idea of this effect on the \ldb. For this reason, we decided to not include this quantity in the cumulative uncertainty computed in Sect. \ref{sec:global}, with the warning that a possible uncertainty source has been neglected. 

\subsection{Outer \bcs}
\label{sec:tau}
To solve the differential equations which describe stellar interiors, suitable outer \bcs{} are required. The common approach consists in specifying pressure $P(\tau_\mathrm{ph})$ and temperature $T(\tau_\mathrm{ph})$ provided by a detailed atmospheric model at a given optical depth $\tau_\mathrm{ph}$ (the matching point between atmosphere and interior). To estimate the effect of \bcs{} on \ldb{} age, two aspects must be considered: (a) the adopted atmospheric model and (b) the choice of $\tau_\mathrm{ph}$. 

The adoption of a detailed non-grey atmospheric structure becomes important when the star has a thick convective envelope and the temperature profile is sensitive to the heat transport in the thin atmospheric layers \citep[see e.g.][]{auman69,dorman89,allard97,baraffe98,montalban04}. 

Unfortunately, detailed atmospheric model tables do not contain the related uncertainties. Given such a situation, one possibility to address the impact of the adopted \bcs{} on \ldb{} age is to compute models with \bcs{} obtained from different atmospheric calculations. In the following we analysed the effect of the adoption of two detailed non-grey atmospheric tables, namely the \citet[][\texttt{BH05}]{brott05}, our reference, and the \citet[][\texttt{AHF11}]{allard11}, computed by means of the same hydrostatic atmospheric code \citep[\textsc{PHOENIX};][]{hauschildt99}. In both cases we used $\tau_\mathrm{ph} = 10$. We computed also models with the grey \citet[][\texttt{KS66}]{krishna66} $T-\tau$ profile, and $\tau_\mathrm{ph}=2/3$. We emphasize that the last set should be considered as an extreme case, since it is well known that the adoption of grey atmospheric model is only a rough approximation of the atmospheric structure of extremely cold low-mass stars \citep[see e.g.][and references therein]{baraffe95}.

A further complication is given by the fact that the \texttt{AHF11} tables for low-mass stars are available for only one metallicity ($Z = 0.0134$). We decided to perform the comparison at a fixed $Z$, so even the \texttt{BH05} (our reference) and the \texttt{KS66} models have been computed for such a metallicity, which, only in this case, is different from the reference one. 

\begin{figure}
	\centering
	\includegraphics[width=\columnwidth]{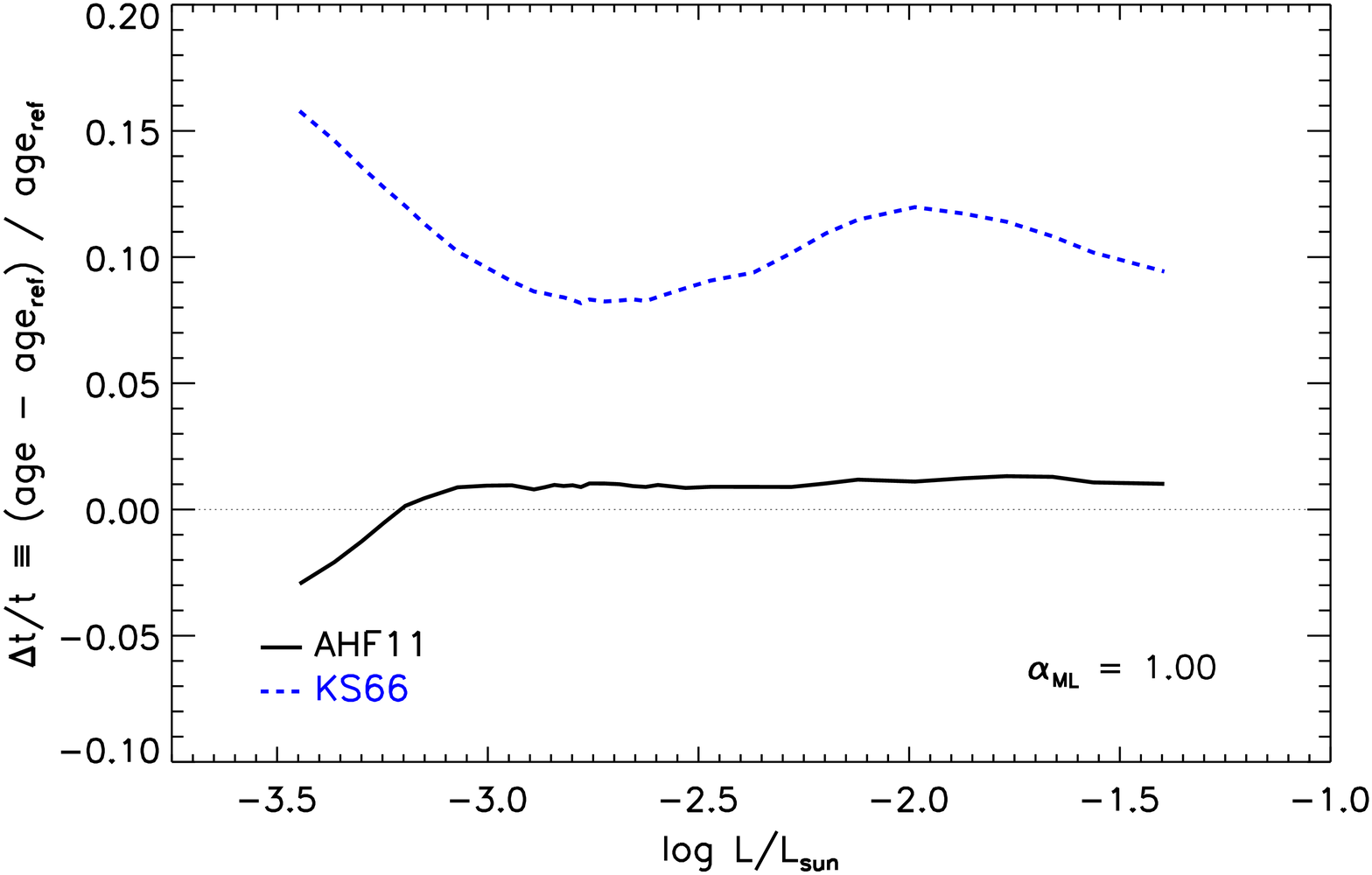}
	\includegraphics[width=\columnwidth]{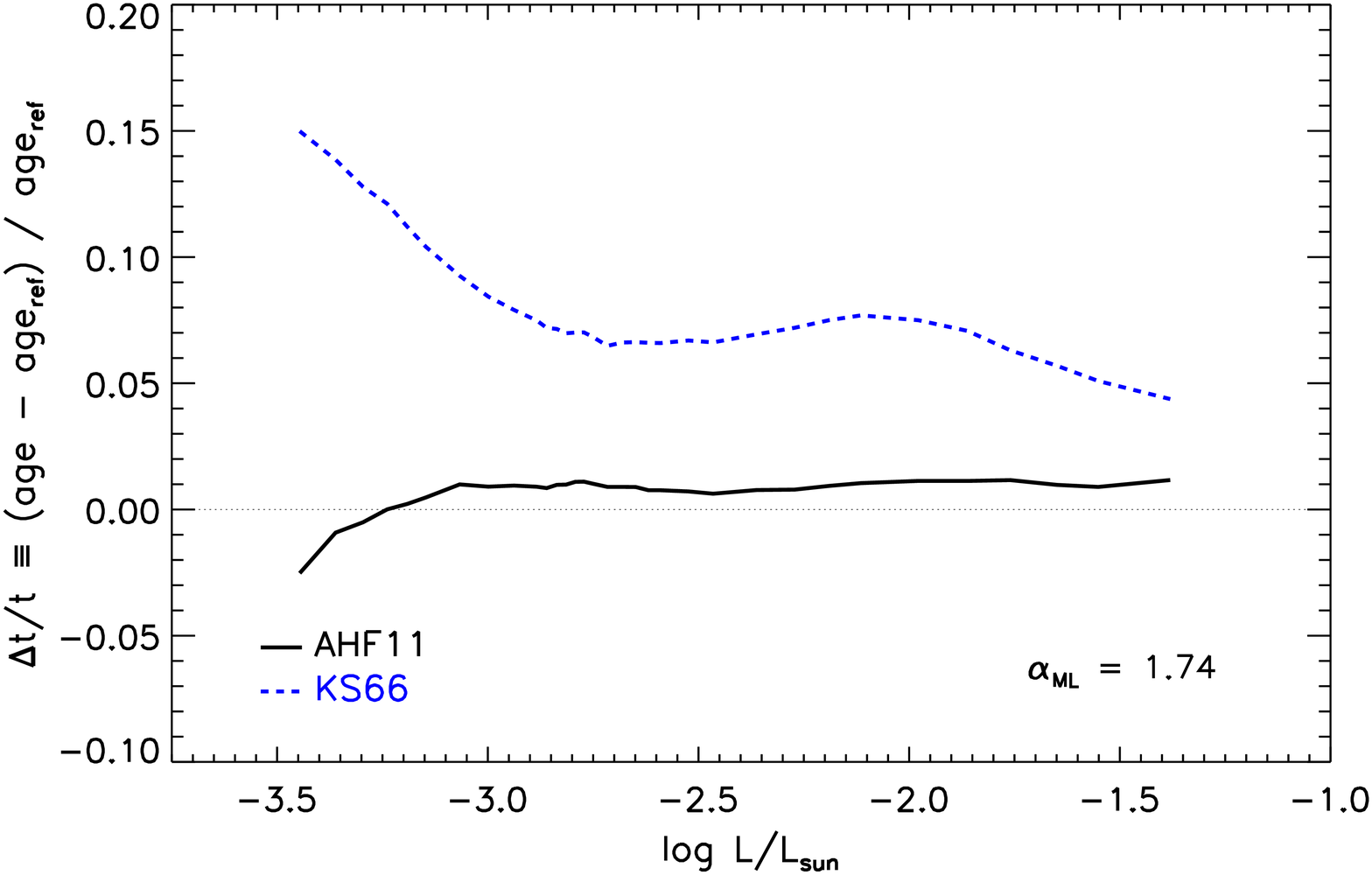}
	\caption{Relative age difference at \ldb{} as a function of luminosity between models with the reference (\texttt{BH05}) and the labelled outer \bcs, for \ml = 1.00 (upper panel) and \ml = 1.74 (bottom panel). As discussed in the text, in this case the metallicity value $Z=0.0134$ has been used instead of the reference one.}
	\label{fig:BCmod}
\end{figure}
Fig. \ref{fig:BCmod} shows the relative age difference versus luminosity due to the adoption of the quoted atmospheric models, for two \ml{} values. As expected given the similarity of the two atmospheric models, the effect of using \texttt{AHF11} in place of \texttt{BH05} is relatively small, being of the order of $1\%$ over a large interval of luminosity and reaching a maximum value of about $3\%$ only at very low luminosities. Note also that, in this case, the relative \ldb{} age differences are essentially the same using \ml{}=1.00 (\emph{upper panel}) or the solar-calibrated value \ml{}=1.74 (\emph{bottom panel}). 

A much stronger effect is due to the adoption of the grey \texttt{KS66}, which leads to relative age differences as large as $8$-$15\%$ for \ml{}=1.00. These differences slightly reduce in the case of \ml{}=1.74 towards the upper bound of the luminosity for those masses where superadiabatic convection occurs in a progressively larger and larger region. 
 
Besides the chosen atmospheric model, \citet{montalban04} showed that the choice of $\tau_\mathrm{ph}$ is important when atmospheric and interior computations do not use the same input physics/prescriptions. In this sense, $\tau_\mathrm{ph}$ is another free-parameter that usually lies in the range [2/3, 100] \citep[see e.g. Table 2 in][]{tognelli11}. 

\begin{figure}
	\centering
	\includegraphics[width=\columnwidth]{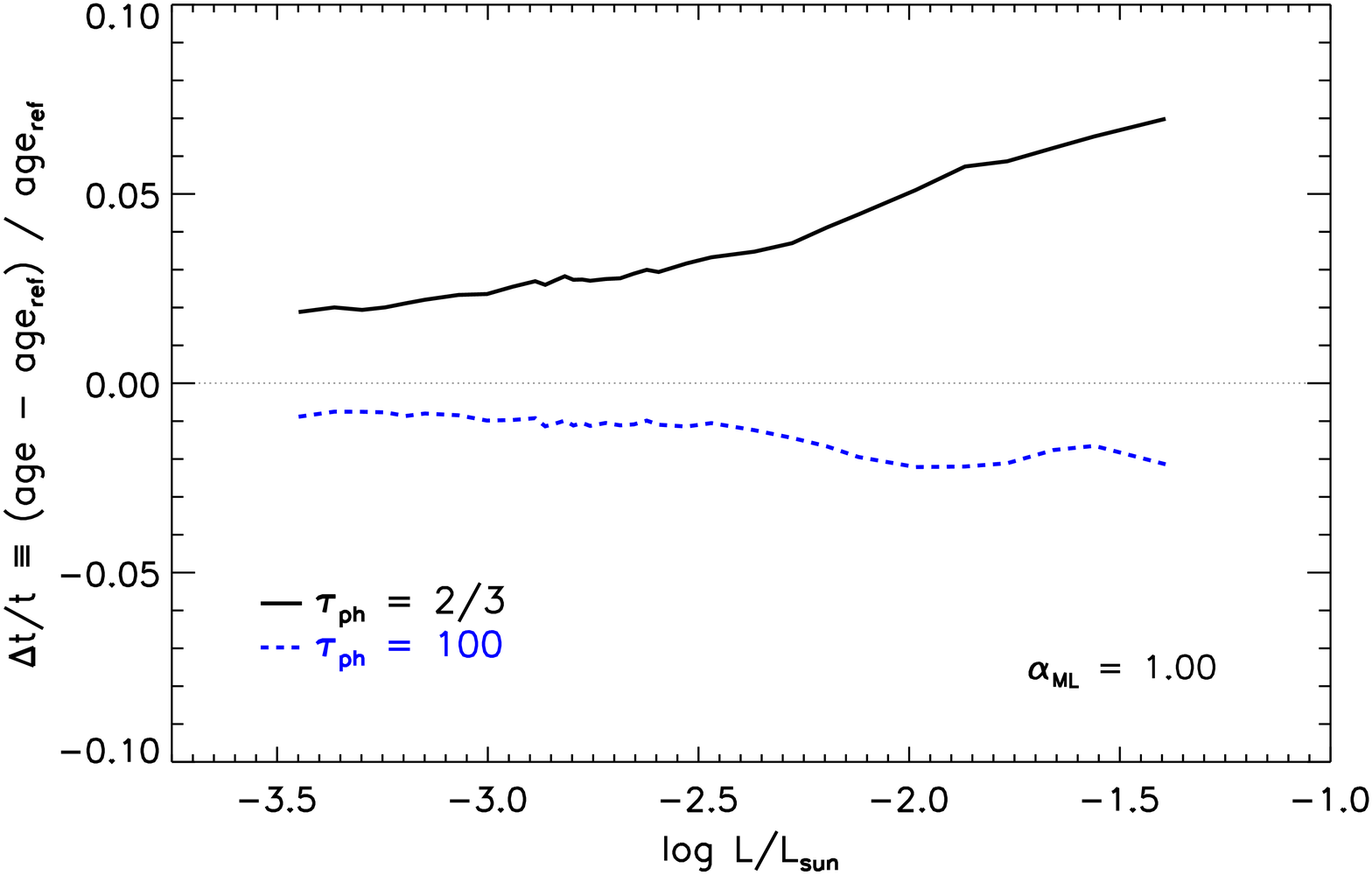}
	\includegraphics[width=\columnwidth]{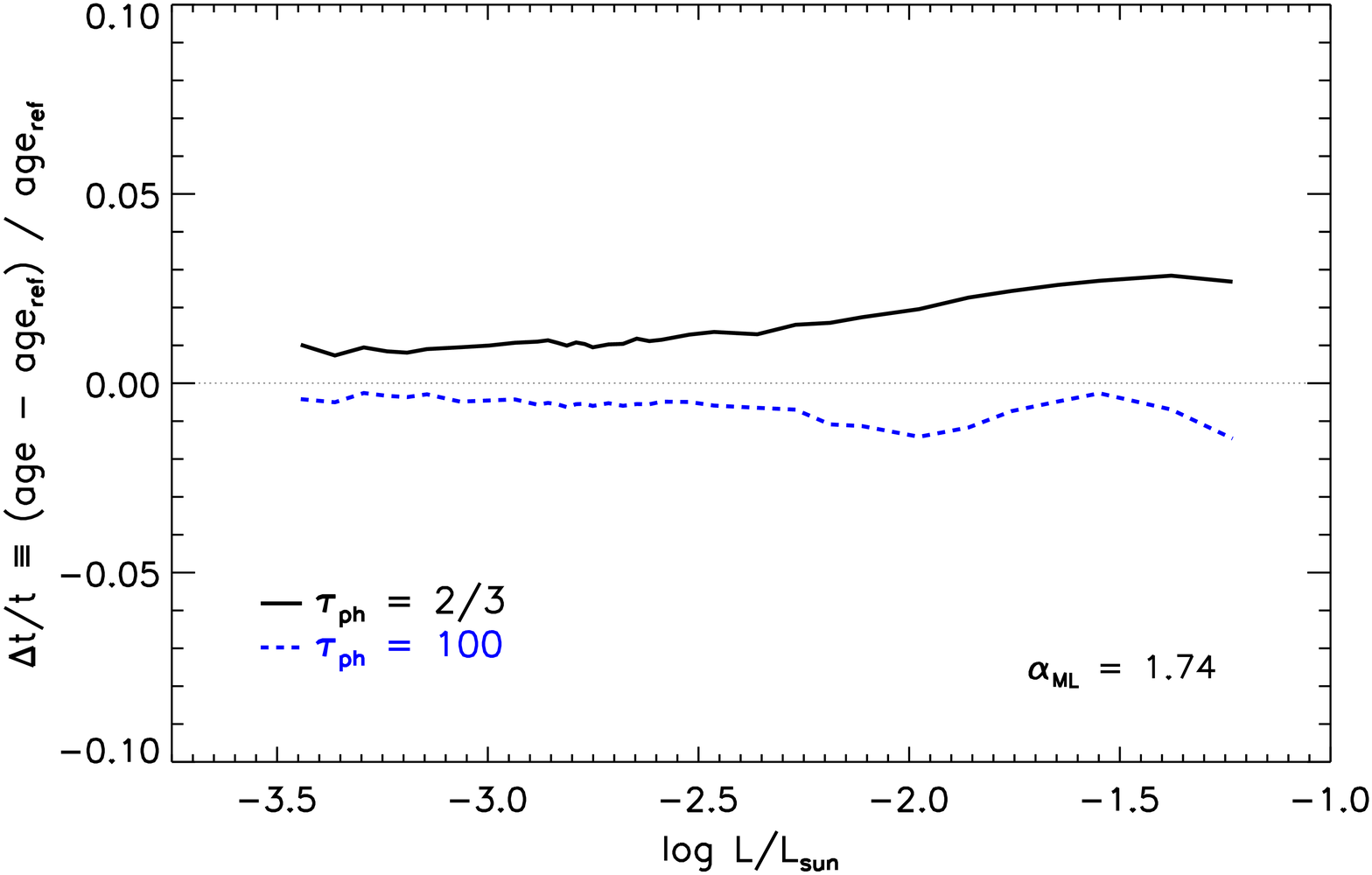}
	\caption{Relative age difference at \ldb{} as a function of luminosity between the reference set of models ($\tau_\mathrm{ph}=10$) and the sets with the labelled $\tau_\mathrm{ph}$ values, for \ml = 1.00 (upper panel) and \ml = 1.74 (bottom panel).}
	\label{fig:tau}
\end{figure}
Fig. \ref{fig:tau} shows the relative age difference versus luminosity obtained varying $\tau_\mathrm{ph}$ and keeping fixed the atmospheric model, i.e. \texttt{BH05}, for two mixing length parameter values. Models are computed for $\tau_\mathrm{ph} = 2/3$, $10$ (reference), and $100$. Passing from $\tau_\mathrm{ph}=$ 10 to 100 the age at the \ldb{} decreases whereas from $\tau_\mathrm{ph}=$ 10 to 2/3 it increases; the largest difference occurs when $\tau_\mathrm{ph}=2/3$ is adopted. The extent of the relative age difference depends on the used mixing length parameter, the effect for \ml=1.00 being roughly twice that for \ml=1.74. Note that all the atmospheric models adopt the  same mixing length value (i.e. $\alpha_\mathrm{ML,atm}=2.00$, for $\tau \lt \tau_\mathrm{ph}$). We are forced to keep it fixed because we do not have the code used to compute the atmospheric structure. 

In convective regions, the temperature gradient becomes progressively more sensitive to the mixing length parameter as superadiabaticity gets larger. The models with $\tau_\mathrm{ph}=2/3$ have a superadiabatic zone in the interior ($\tau \ge \tau_\mathrm{ph}$), where we can actually change \ml, larger than the model with $\tau_\mathrm{ph}=100$. The latter set of models has consequently a reduced sensitivity to the adopted mixing length parameter in the interior.

Given the difficulty in estimating the uncertainty in the adopted \bcs, for the computation of the cumulative error stripe in \ldb{} age in Sect. \ref{sec:global}, we took into account only the effect of changing $\tau_\mathrm{ph}$.

\subsection{Radiative opacity}
Although radiative opacity is one of the main ingredients in stellar computations, the current generation of tables do not contain any indication about the uncertainty. As a first step, to give an estimate of the opacity uncertainty propagation on the \ldb{} age, we analysed the effect of an uncertainty in the Rosseland coefficients ($\overline{\kappa}_\mathrm{rad}$) of $\pm 5\%$ in the whole structure, as done in \citet{valle13a}.

In convective regions, the actual temperature gradient depends on the radiative opacity only in superadiabatic zones, while elsewhere it is mainly determined by the \eos. In the inner region of a fully convective star, given the high density, the temperature gradient is essentially the adiabatic one, while it becomes progressively more and more superadiabatic moving towards the surface, where the density drops. It is thus clear that in the interior of stars, the change of the opacity coefficients has a negligible effect on the structure, whereas the effect increases towards the surface. 

For models computed with detailed non-grey \bcs, we can not modify the atmospheric opacity because we do not have the atmospheric code, thus preventing the possibility to consistently check the effect of $\overline{\kappa}_\mathrm{rad}$. Moreover, for the selected mass range in the interior ($\tau \ge \tau_\mathrm{ph}$), where we can actually modify the opacity coefficients, the density is high enough to guarantee an almost adiabatic convection. For this set of non-grey models we verified that the \ldb{} age is not affected by a variation of the radiative opacity coefficients in the interiors (for both $\tau_\mathrm{ph} = 2/3$ and 10).

A way in which we can estimate the effect of an opacity variation extended also to the atmosphere is to compute stellar models adopting grey outer \bcs. In this case we can actually calculate the atmospheric structure consistently with the $\overline{\kappa}_\mathrm{rad}$ variation used in the interior. Although we are aware that the adoption of a grey \bc{} is not the best choice for the mass range we are dealing with, such an approach allowed us to give an estimate of the effect on \ldb{} age of an opacity variation in the whole structure, atmosphere included.
	
\begin{figure}
	\centering
	\includegraphics[width=\columnwidth]{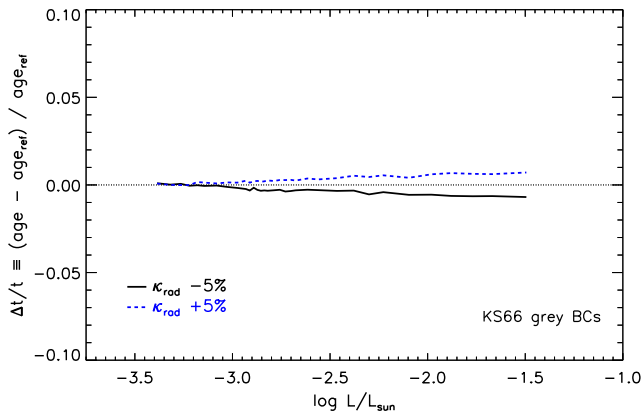}
	\caption{Relative age difference at \ldb{} as a function of luminosity between models computed with the reference and the $\pm5\%$ perturbed radiative opacity. Grey outer \bcs{} are adopted instead of the reference ones.}
	\label{fig:ks66_kappa}
\end{figure}
Fig. \ref{fig:ks66_kappa} shows the relative age difference versus luminosity caused by a variation of $\pm 5\%$ of the radiative opacity coefficients. Due to the progressively more extended superadiabatic zone in the outermost layers at increasing stellar mass, the relative age uncertainty increases with the luminosity from negligible values at the faint end to about $1\%$ at the bright end. At a given luminosity, the lower the opacity and the larger the central temperature and, consequently, the lower the \ldb{} age. 

\begin{figure}
	\centering
	\includegraphics[width=\columnwidth]{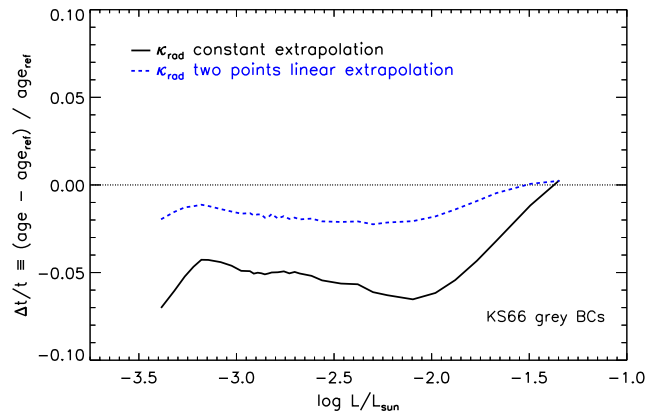}
	\caption{Relative age difference at \ldb{} as a function of luminosity between the reference models and those computed with radiative opacity tables extrapolated with different procedures. Grey outer boundary conditions are adopted instead of the reference ones.}
	\label{fig:ks66_extra}
\end{figure}
A further uncertainty source related to radiative opacity is worth to be discussed. The radiative opacity tables do not cover the entire temperature$-$density plane required for computing \pms{} evolution for masses in the range [0.06, 0.4] M$_\odot$ and an extrapolation to higher densities is needed. As in \citet{tognelli11}, in order to evaluate the impact of the extrapolation on the \ldb{} age, we analysed three different extrapolation techniques: constant, linear from the last two points, and a linear fit from the last four points (our reference). As in the previous cases, we cannot properly evaluate the effect of changing the extrapolation on the \ldb{} age for non-grey models, because we can not modify the atmospheric models. The only thing that we can analyse for non-grey models is the effect of changing the extrapolation in the interiors. As expected the effect of such a change is negligible. 

However, to have an idea of the extrapolation impact we computed three additional sets of models with grey \bcs{} where we can actually change the extrapolated opacities also in the atmosphere. Fig. \ref{fig:ks66_extra} shows the relative \ldb{} age difference between models computed with the reference, i.e. four point linear, and the constant and two points linear radiative opacity extrapolations. The last two extrapolations lead to underestimate the \ldb{} age with respect to the reference one; however, such an effect vanishes at high luminosities. The constant extrapolation produces an effect which is about three times larger than the linear one.

\begin{figure}
	\centering
	\includegraphics[width=\columnwidth]{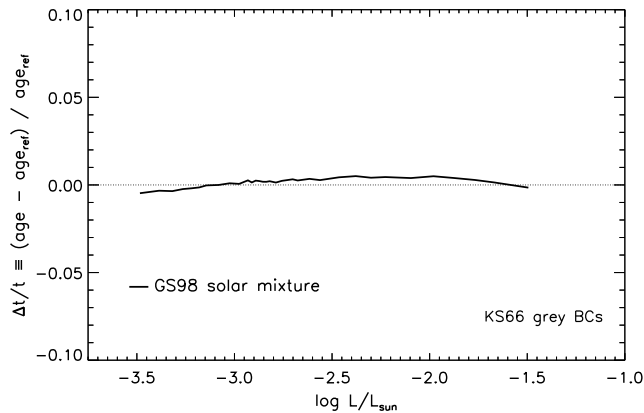}
	\caption{Relative age difference at \ldb{} as a function of luminosity between models computed with radiative opacity adopting \texttt{AS09} (reference) and \texttt{GS98} solar mixtures. Grey outer \bcs{} are adopted instead of the reference ones.}
	\label{fig:ks66_mix}
\end{figure}
A final uncertainty source related to radiative opacity is the heavy elements distribution adopted to compute Rosseland coefficients. To analyse the effect on the \ldb{} age of changing the heavy element mixture at a fixed total metallicity $Z$, we computed models adopting the \citet[][\texttt{GS98}]{grevesse98} solar mixture. As in the previous cases, we discuss only the effect on grey models since non-grey ones are completely unaffected also by the mixture change in the opacity tables used to compute the interiors, the mixture adopted by \texttt{BH05} atmospheric models being fixed. Fig. \ref{fig:ks66_mix} shows the relative \ldb{} age difference between models adopting the reference (\texttt{AS09}) and the \texttt{GS98} solar mixture and computed with the grey \bcs. Depending on the mass, the stellar structure crosses a region in the density$-$temperature plane where the opacity coefficients are particularly sensitive to different elements \citep[see e.g.][]{sestito06}. Thus, the opacity variation due to the adopted mixture is a complex function of temperature and density. However, as clearly visible in Fig. \ref{fig:ks66_mix}, the effect induced by the adopted mixture is very small, being always in the range [$-0.5\%$, $+0.5\%$].

We computed sets of models with the reference and perturbed radiative opacity coefficients ($\overline{\kappa}_\mathrm{rad} \pm5\%$, different high-density extrapolations, different heavy elements mixtures) also for \ml{}=1.74. The effect of an opacity variation on \ldb{} age is completely insensitive to a mixing length change.

Summarizing the results of this section, we showed that changing the radiative opacity only in the interiors does not affect the \ldb{} age. On the contrary, a sizeable effect is expected when the radiative opacity variation is extended to the atmosphere. However, since we were able to compute such an effect only in the case of grey \bcs, we preferred to neglect the opacity contribution in the computation of the cumulative uncertainty in the \ldb{} age (see Sect. \ref{sec:global}), being aware that a possible uncertainty source is missing in the cumulative error estimate.

\subsection{Equation of state}
The structure of fully convective \pms{} stars depends on the \eos{} \citep[see e.g.,][]{mazzitelli89,dantona93,tognelli11}. 

As for radiative opacity, the most commonly adopted \eos{} tables do not contain the uncertainty associated with the thermodynamic quantities (i.e. density, specific heat, adiabatic gradient etc.). Moreover, these quantities are strictly correlated among each other, thus preventing to follow a similar approach to estimate the uncertainty propagation into \ldb{} age as that used for opacity. For this reason, to roughly estimate the dependence of the \ldb{} age on the \eos, in addition to our reference set of models (\textsc{OPAL06}+\textsc{SCVH95 EOS}), we computed three additional sets adopting different \eos{} tables. We used the sole \textsc{OPAL06} \eos{}, the \textsc{FreeEOS} \citep[][in the \texttt{EOS1} configuration]{irwin08}, and the sole \textsc{SCVH95} \eos. Since both the \textsc{FreeEOS} and \textsc{OPAL06} \eos{} do not cover the whole temperature$-$density plane suitable for computing the entire mass range [0.06, 0.4] M$_\odot$ studied in this paper, we limit the comparison to the common mass interval. 

\begin{figure}
	\centering
	\includegraphics[width=\columnwidth]{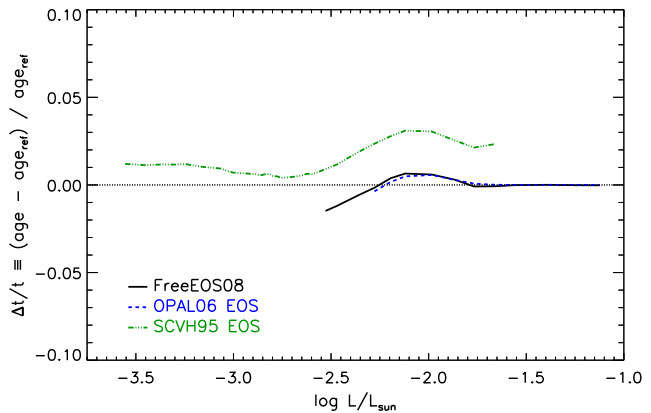}
	\caption{Relative age difference at \ldb{} as a function of luminosity between the reference set of models (\textsc{OPAL06+SCVH95}) and the sets with the labelled \eos.}
	\label{fig:eos}
\end{figure}
Fig. \ref{fig:eos} shows the relative age difference versus luminosity due to the adoption of the quoted \eos. Both \textsc{FreeEOS} and \textsc{OPAL06} induce age differences less than $\pm 1\%$. A larger effect 1-3$\%$ is caused by the \textsc{SCVH95} \eos. \citet{burke04} already studied the effect on \ldb{} of the \eos, but they used tables different from those adopted here, thus preventing a detailed comparison. However, their results are, at least qualitatively, in agreement with ours. We also verified that the adoption of \ml{}=1.74 in the reference and perturbed \eos{} models leads to the same relative \ldb{} age differences.

Due to the difficulty to properly estimate the uncertainty propagation of the adopted \eos, we did not take such an effect into account for the computation of the cumulative error bars in \ldb{} age in Sect. \ref{sec:global}.

\subsection{Mixing length efficiency}
Although commonly adopted, the mixing length theory is not a fully consistent treatment of the superadiabatic convection and, consequently, it introduces a relevant uncertainty source in computation of stellar structures with extended convective envelopes. The commonly adopted solar calibration of the mixing length parameter \ml{} does not in principle guarantee a correct convective efficiency for stars, as in our case, of mass and/or evolutionary phase different from that of the Sun. In order to check the effect of changing the mixing length value on the \ldb{} age, we computed two sets of models with \ml = 1.74 (our solar calibrated) and the other with \ml = 1.00, as suggested by some authors for \pms{} stars \citep[see][]{stassun08,gennaro12,tognelli12,somers14}

\begin{figure}
	\centering
	\includegraphics[width=\columnwidth]{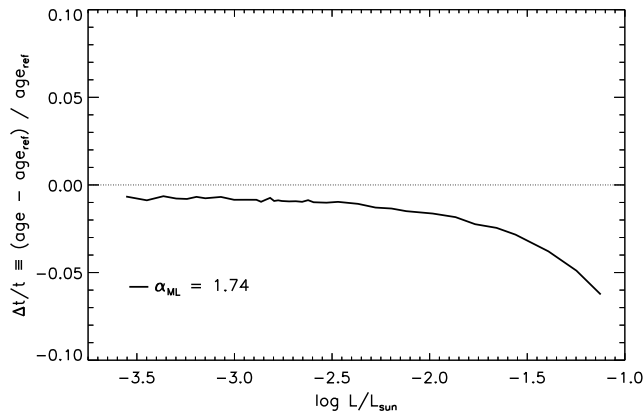}
	\caption{Relative age difference at \ldb{} as a function of luminosity between sets computed with two different mixing length parameter values, \ml = 1.00 (reference) and \ml = 1.74 (solar calibrated).}
	\label{fig:ml}
\end{figure}
Fig. \ref{fig:ml} shows the relative age difference versus luminosity due to the adoption of the quoted mixing length parameters. The effect of changing \ml{} from 1.00 to 1.74 is to increase the central temperature, hence to increase the efficiency of \libur{}, and to reduce the age at the \ldb. The relative age difference of the \ldb{} is of the order of about $-1\%$ and $-6\%$ for the lowest and, respectively, highest luminosities. The effect is lower at very low luminosity (very low mass stars) as a result of the thinner superadiabatic region. Such a result  is in agreement with that shown by \citet{burke04}.

\section{Individual chemical composition uncertainties}
\label{sec:var_indi_chm}
Besides the input physics, \pms{} stellar models depend also on the adopted chemical composition, namely the initial metallicity $Z$, helium $Y$ and deuterium $X_\mathrm{d}$ abundances. Thus, when comparing observations and theoretical models, the proper chemical composition should be adopted. However, quite often only independent measurements of the current photospheric [Fe/H] of the stellar cluster/group are available, whereas direct estimates of $Y$, $X_\mathrm{d}$, and metal abundances are missing. To overcome such a lack various assumptions are needed. 

The [Fe/H] value is converted into the total metallicity $Z$ adopting, for Population I stars, a \textit{solar-scaled} distribution of metals, and assuming an initial helium abundance. Stellar modellers usually compute initial $Y$ using the following linear relation:
\begin{equation}
Y = Y_{\mathrm{P}} + \frac{\Delta Y}{\Delta Z} Z,
\label{eq:elio}
\end{equation}
where $Y_{\mathrm{P}}$ is the primordial helium abundance and $\Delta Y/\Delta Z$ is the helium-to-metal enrichment ratio. The proper initial metallicity $Z$ to be adopted in stellar models follows from:
\begin{equation}
Z = \frac{(1-Y_{\mathrm{P}})(Z/X)_\odot }{10^{-[\mathrm{Fe/H}]}+(1+\Delta Y/\Delta Z)(Z/X)_\odot},
\label{eq:metal}
\end{equation}
where $(Z/X)_\odot$ is the current solar photospheric metal-to-hydrogen ratio. For the reference set of models we adopted $Y_\mathrm{P}=0.2485 \pm 0.0008$ \citep{cyburt04}, $\Delta Y/\Delta Z = 2$ \citep{casagrande07}, and $(Z/X)_\odot = 0.0181$ \citep{asplund09}.

All the quantities in equations (\ref{eq:elio}) and (\ref{eq:metal}) are known with an error, shown in Table \ref{tab:err_chm}, which directly propagates into the final $Y$ and $Z$ values. Note that these errors are intended to represent the edges of the variability region, rather than confidence intervals. Similarly to what was done in the previous sections, as a first step we computed perturbed models by varying a single parameter ($\Delta Y/ \Delta Z$, [Fe/H], and $(Z/X)_\odot$) at a time keeping all the others fixed to the reference value. In the following, we did not take into account the uncertainty in $Y_\mathrm{P}$, being negligible. 

To the best of our knowledge, this is the first detailed analysis of the chemical composition uncertainty in the \ldb{} age estimates.

\begin{table}
\centering
\caption{Chemical composition parameters varied in the computation of perturbed stellar models and their assumed uncertainty (see text). The flag `yes' in the last column specifies the quantities taken into account in the cumulative uncertainty calculation (see Sect. \ref{sec:global}).}
\label{tab:err_chm}
\begin{tabular}{lcc}
\hline
Quantity & Error & Global\\
\hline
\hline
$[$Fe/H$]$ & $\pm 0.1$dex & Yes\\
$\Delta Y/\Delta Z$ & $\pm 1$& Yes\\
$(Z/X)_\odot$ & $\pm 15\%$ & Yes\\
$X_\mathrm{d}$ & $\pm 1\times 10^{-5}$ & Yes\\
\hline
\end{tabular}
\end{table}

\subsection{Initial helium abundance}
Following equation (\ref{eq:elio}), the initial helium abundance adopted in stellar models strongly depends on the helium-to-metal enrichment ratio, which is poorly constrained by observations \citep[see e.g.][]{gennaro10}. We adopted an error on $\Delta Y/\Delta Z $ of $\pm 1$ \citep{casagrande07}, and we kept fixed the value of $Y_\mathrm{P}$. In order to quantify the effect of the quoted uncertainty on the \ldb{} age, we computed two sets of models with the reference $Z=0.0130$ and $X_\mathrm{d} = 2\times10^{-5}$, and two different values of $Y$, namely $0.261$ and $0.287$, which correspond to $\Delta Y/\Delta Z = 1$ and $3$.

\begin{figure}
	\centering
	\includegraphics[width=\columnwidth]{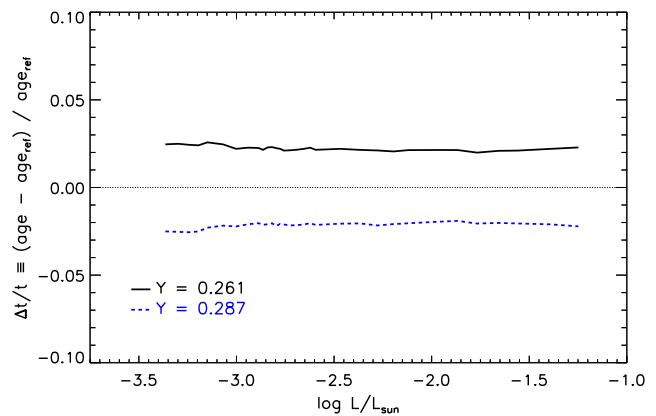}
	\caption{Relative age difference at \ldb{} as a function of luminosity between the reference set of models ($Y=0.274$) and the sets with the labelled initial helium abundance $Y$.}
	\label{fig:y}
\end{figure}
Fig. \ref{fig:y} shows the relative age difference versus luminosity due to the adoption of the quoted initial helium abundances. The relative age uncertainty is roughly $\pm2\%$ over the whole mass range. The larger the helium abundance, the higher the central temperature, and the earlier the lithium depletion at a given luminosity. The results do not change if \ml{}=1.74 is used instead of \ml{}=1.00.

\subsection{Initial metallicity}
In order to analyse the effect of the metallicity on the \ldb{} age, we computed two sets of models with the reference $Y=0.274$ and $X_\mathrm{d} = 2\times10^{-5}$, and two different metallicity values, $Z=0.0105$ and $Z=0.0155$, which roughly correspond to the typical observational error in [Fe/H], namely $\pm 0.1$ dex.

\begin{figure}
	\centering
	\includegraphics[width=\columnwidth]{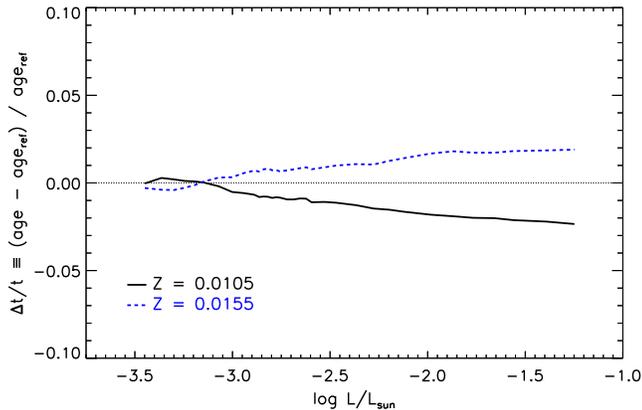}
	\caption{Relative age difference at \ldb{} as a function of luminosity between the reference set of models ($Z=0.0130$) and the sets with the labelled initial metallicity $Z$.}
	\label{fig:z}
\end{figure}
Fig. \ref{fig:z} shows the relative age difference versus luminosity due to the adoption of the quoted initial metallicities. The effect is almost negligible at faint end of our range and it increases up to about $\pm2\%$ at higher luminosities. Such a behaviour is qualitatively similar to that shown in Fig. \ref{fig:ks66_kappa}, as a variation in metallicity mainly translates into a variation in the radiative opacity. This is essentially the consequence of the metallicity dependence of the outer \bcs. We verified that varying the metallicity only in the interiors, keeping fixed the \bcs{}, produces a very small ($\la 0.5\%$) and constant effect on the \ldb. In this case at a fixed luminosity, a larger $Z$ leads to a larger mass (higher central temperatures) and consequently to a lower \ldb{} age. Moreover, we verified that decreasing the luminosity, the pressure and temperature at the base of the atmosphere of models at the \ldb{} get progressively less and less sensitive to the metallicity. At higher luminosity, the effect of $Z$ on the \bcs{} becomes dominant and an increase of $Z$ results in a larger \ldb{} age. The comparison between reference and perturbed models computed with \ml{}=1.74 gives the same results.

\subsection{Heavy elements mixture}
The value of $(Z/X)_\odot$ in equation (\ref{eq:metal}) depends on the adopted solar heavy elements mixture. In our reference set of models we used the \citet{asplund09} one, which leads to $(Z/X)_\odot=0.0181$. In order to quantify the impact of varying this quantity on \ldb{} age, we computed two sets of models with the reference values of $\Delta Y/\Delta Z = 2$, [Fe/H] = $+0.0$, and $X_\mathrm{d}=2\times 10^{-5}$, adopting an uncertainty of $\pm15\%$ in $(Z/X)_\odot$ \citep{bahcall04,bahcall05}. This is also roughly representative of the range of values spanned by different, but still largely used, solar mixtures \citep[$(Z/X)_\odot \approx 0.0165$ - $0.0244$;][]{grevesse93,grevesse98,asplund05,caffau08}. 
Hence, using the lower and upper values of $(Z/X)_\odot$ we computed the corresponding models with ($Y$, $Z$) = (0.271, 0.0120) and (0.278, 0.0150).

\begin{figure}
	\centering
	\includegraphics[width=\columnwidth]{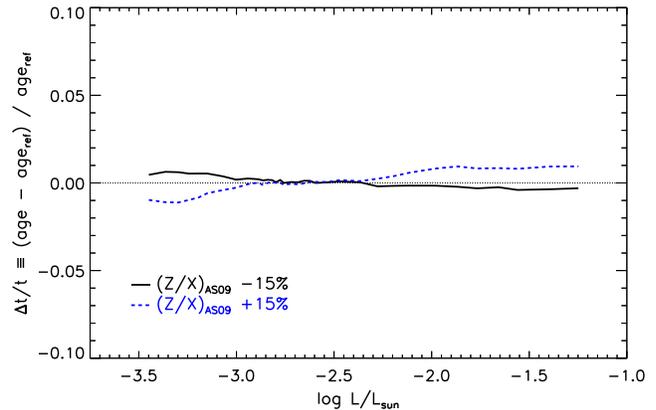}
	\caption{Relative age difference at \ldb{} as a function of luminosity between models adopting the reference ($(Z/X)_\odot=0.0181$) and the labelled $(Z/X)_\odot$ values.}
	\label{fig:mix}
\end{figure}
Fig. \ref{fig:mix} shows the relative age difference versus luminosity due to the adoption of the quoted initial metallicities and helium abundances. The shown behaviour is the consequence of a variation of both $Z$ and $Y$ resulting from a change of $(Z/X)_\odot$ at a fixed [Fe/H] and $\Delta Y/\Delta Z$. Note that, as shown in Figs. \ref{fig:y} and \ref{fig:z}, increasing the luminosity, the effect of a variation on $Z$ is opposite and partially counterbalanced by that of varying $Y$, leading to a total effect lower than $\pm 1\%$. The relative \ldb{} age difference between reference and perturbed models is not affected by changing \ml{} from 1.00 to 1.74. 

\subsection{Initial deuterium abundance}
Although its tiny initial abundance, deuterium plays a relevant role in \pms{} evolution because the energy released during d-burning temporarily slows down the gravitational contraction. In order to evaluate the effect of the initial deuterium abundance uncertainty on the \ldb{} age, we computed two sets of models with $X_\mathrm{d} = 1\times10^{-5}$ and $3\times 10^{-5}$ representative of the current $X_\mathrm{d}$ range in the Galactic disc/local bubble/solar neighbourhood \citep[see e.g.,][]{sembach10}.

\begin{figure}
	\centering
	\includegraphics[width=\columnwidth]{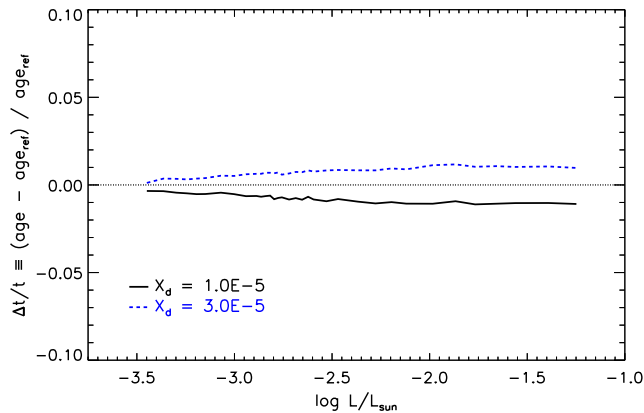}
	\caption{Relative age difference at \ldb{} as a function of luminosity between models adopting the reference ($X_\mathrm{d}=2\times10^{-5}$) and the labelled initial deuterium abundances.}
	\label{fig:xd}
\end{figure}
Fig. \ref{fig:xd} shows the relative age difference versus luminosity due to the adoption of the quoted initial deuterium abundances. The effect is always less than $\pm1\%$. The larger is the $X_\mathrm{d}$, the longer is the d-burning phase, and thus the later is the lithium depletion at a given luminosity. The relative \ldb{} age difference due to the initial deuterium abundance variation does not change if \ml{}=1.74 is used.

\section{Cumulative physical and chemical uncertainties}
\label{sec:global}
In the previous section, we analysed the effect on \ldb{} age of varying a single input physics at a time by keeping fixed all the others, and varying the chemical composition for the reference input physics. However, although commonly adopted \citep{bildsten97,burke04}, such an approach does not take into account possible interactions among the different ingredients of stellar models. In other words, it assumes that non-linear effects due to simultaneous variation of such quantities are absent or negligible. 

For this reason, we decided to follow a more suitable procedure similar to that described in \citet{valle13a}, consisting in the systematic and simultaneous variation of the main input on a fixed grid. More in detail, stellar models and hence the resulting \ldb{} age depend on a vector of parameters $\{p_j\}$ (i.e. input physics and chemical elements abundances, as discussed in the previous sections). Each parameter $p_j$ can assume three values, namely $p_{j,0} - \delta p_j$, $p_{j,0}$, and $p_{j,0} + \delta p_j$, where $p_{j,0}$ represents the reference value and $\delta p_j$ the adopted uncertainty. In order to cover the whole parameter space of the simultaneously varied input, we computed a set of perturbed stellar models for each possible combination of the vector $\{p_j\}$. Such a \emph{distribution-free} technique relies only on the specification of a sensible range of variation for each input and it does not require an explicit specification of the parent distributions of the varied parameters. Thus, it allows us to determine the edges of the variability region rather than a confidence interval. Such a technique is particularly useful when no information about the errors distribution on the analysed quantity is available.

This method is more robust than assuming an a priori independence of the input physics/chemical parameters impact, but it has the serious disadvantage of being much more computationally time consuming than the classic approach. So, in order to avoid a waste of time in useless computations, we did not vary all the input physics studied in Sect. \ref{sec:var_indi_fis}. We did not perturb the d-burning cross-sections (p+$^2$H, $^2$H+$^2$H), since we showed that they have a completely negligible effect on the \ldb{} age. Moreover, as discussed in the previous section, we did not vary the \eos, radiative opacity and atmospheric models since we can not properly take into account their effects. The list of the input physics allowed to vary is shown in the last column of Tables \ref{tab:err_fis} and \ref{tab:err_chm}. Note that the  perturbed stellar models can not be ruled out by the current observational constraints. In fact, we checked that the relative radius variation due to the perturbed input physics at fixed mass in the very low mass regime is smaller than 1\%.

\begin{table}
\centering
\caption{Pairs of ($Y$,$Z$) values adopted for the computations of the models with the perturbed chemical composition used for constructing the cumulative error stripe (see text).}
\label{tab:chimica_usata}
\begin{tabular}{cccc}
\hline
($Y$, $Z$) & [Fe/H] & $\Delta Y / \Delta Z$ & $(Z/X)_\odot$\\
\hline
\hline
$Y=0.2740$ & \multirow{2}{*}{$+0.0$} & \multirow{2}{*}{$2$} & \multirow{2}{*}{\texttt{AS09}}\\
$Z=0.0130$  & \\
\hline
$Y=0.2790$ & \multirow{2}{*}{$-0.1$} & \multirow{2}{*}{$3$} & \multirow{2}{*}{\texttt{AS09}}\\
$Z=0.0100$  & \\
\hline
$Y=0.2650$ & \multirow{2}{*}{$+0.1$} & \multirow{2}{*}{$1$} & \multirow{2}{*}{\texttt{AS09}}\\
$Z=0.0160$  & \\
\hline
$Y=0.2750$ & \multirow{2}{*}{$-0.1$} & \multirow{2}{*}{$3$} & \multirow{2}{*}{\texttt{AS09}$-15\%$}\\
$Z=0.0090$  & \\
\hline
$Y=0.2670$ & \multirow{2}{*}{$+0.1$} & \multirow{2}{*}{$1$} & \multirow{2}{*}{\texttt{AS09}$+15\%$}\\
$Z=0.0190$  & \\
\hline
$Y=0.2570$ & \multirow{2}{*}{$-0.1$} & \multirow{2}{*}{$1$} & \multirow{2}{*}{\texttt{AS09}$-15\%$}\\
$Z=0.0090$  & \\
\hline
$Y=0.3020$ & \multirow{2}{*}{$+0.1$} & \multirow{2}{*}{$3$} & \multirow{2}{*}{\texttt{AS09}$+15\%$}\\
$Z=0.0190$  & \\
\hline
\end{tabular}
\end{table}
\begin{table}
\centering
\caption{Pairs of ($Y$,$Z$) values adopted for computing models with the individual perturbation of [Fe/H], $\Delta Y / \Delta Z$, and $(Z/X)_\odot$ (see text).}
\label{tab:chimica_usata_indi}
\begin{tabular}{cccc}
\hline
($Y$, $Z$) & [Fe/H] & $\Delta Y / \Delta Z$ & $(Z/X)_\odot$\\
\hline
\hline
$Y=0.2710$ & \multirow{2}{*}{$+0.0$} & \multirow{2}{*}{$2$} & \multirow{2}{*}{\texttt{AS09}$-15\%$}\\
$Z=0.0120$  & \\
\hline
$Y=0.2780$ & \multirow{2}{*}{$+0.0$} & \multirow{2}{*}{$2$} & \multirow{2}{*}{\texttt{AS09}$+15\%$}\\
$Z=0.0150$  & \\
\hline
$Y=0.2620$ & \multirow{2}{*}{$+0.0$} & \multirow{2}{*}{$1$} & \multirow{2}{*}{\texttt{AS09}}\\
$Z=0.0130$  & \\
\hline
$Y=0.2870$ & \multirow{2}{*}{$+0.0$} & \multirow{2}{*}{$3$} & \multirow{2}{*}{\texttt{AS09}}\\
$Z=0.0130$  & \\
\hline
$Y=0.2690$ & \multirow{2}{*}{$-0.1$} & \multirow{2}{*}{$2$} & \multirow{2}{*}{\texttt{AS09}}\\
$Z=0.0100$  & \\
\hline
$Y=0.2810$ & \multirow{2}{*}{$+0.1$} & \multirow{2}{*}{$2$} & \multirow{2}{*}{\texttt{AS09}}\\
$Z=0.0160$  & \\
\hline
\end{tabular}
\end{table}
Regarding the chemical composition, the number of ($Y$, $Z$) pairs resulting from all the possible combinations of [Fe/H]$\pm 0.1$ dex, $\Delta Y / \Delta Z \pm 1$, and $(Z/X)_\odot \pm 15\%$ values is 27 ($=3^3$). However, it is not needed to calculate set of models for each ($Y$, $Z$) pair. Indeed, being interested only in the edges of the variability region, it is enough to compute stellar models for the 7 $(Y,\,Z)$ pairs listed in Table \ref{tab:chimica_usata}, which correspond to the reference set and to the sets with the extreme combinations of helium abundance and/or metallicity.

In summary, we have adopted for the cumulative error evaluation seven pairs of ($Y$, $Z$), three initial deuterium abundances $X_\mathrm{d}$, two mixing length values \ml{}, three $\tau_\mathrm{ph}$ and three $^7$Li(p,$\alpha$)$\alpha$ cross-section values, for a total of 378 ($=7\times 3\times 2 \times 3 \times 3$) sets of models. For each set we computed a grid of 31 stellar masses, for a total of 11718 stellar tracks.

From the perturbed stellar models, we then obtained the \emph{cumulative error stripe}, following the algorithm of constructing the convex hull described in appendix A in \citet{valle13a}. Here we briefly summarize such a procedure. Each set of perturbed stellar models provides an \ldb{} curve in the age$-$luminosity plane. Thus, at a given luminosity an ensemble of perturbed \ldb{} ages is available. The convex hull of such an ensemble provides the error in the predicted \ldb{} age for the chosen luminosity\footnote{In this case the convex hull coincides with the maximum and minimum values of the \ldb{} age at the chosen luminosity.}. The error stripe is then obtained by changing the luminosity.

\begin{figure}
	\centering
	\includegraphics[width=\columnwidth]{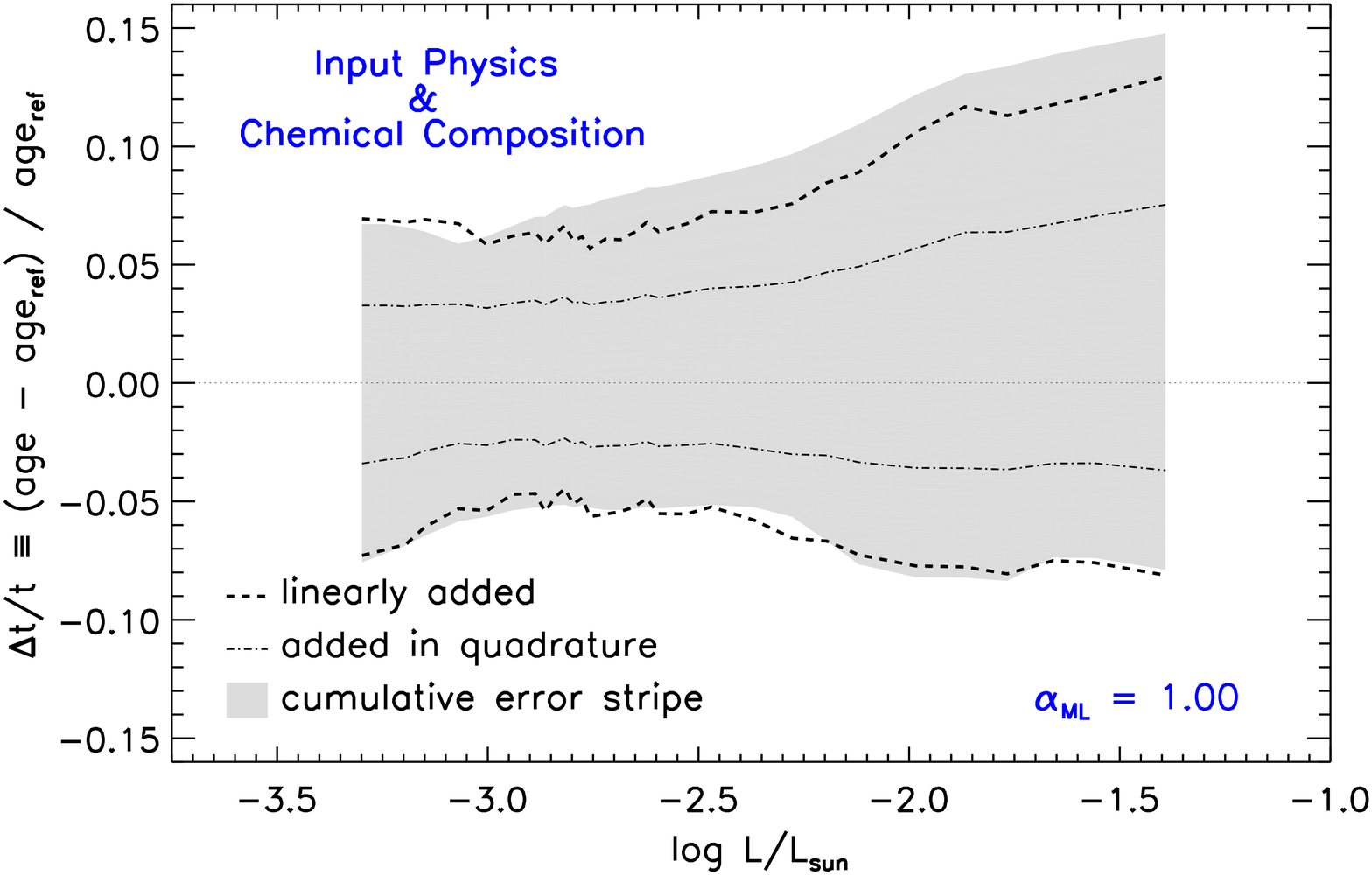}
	\includegraphics[width=\columnwidth]{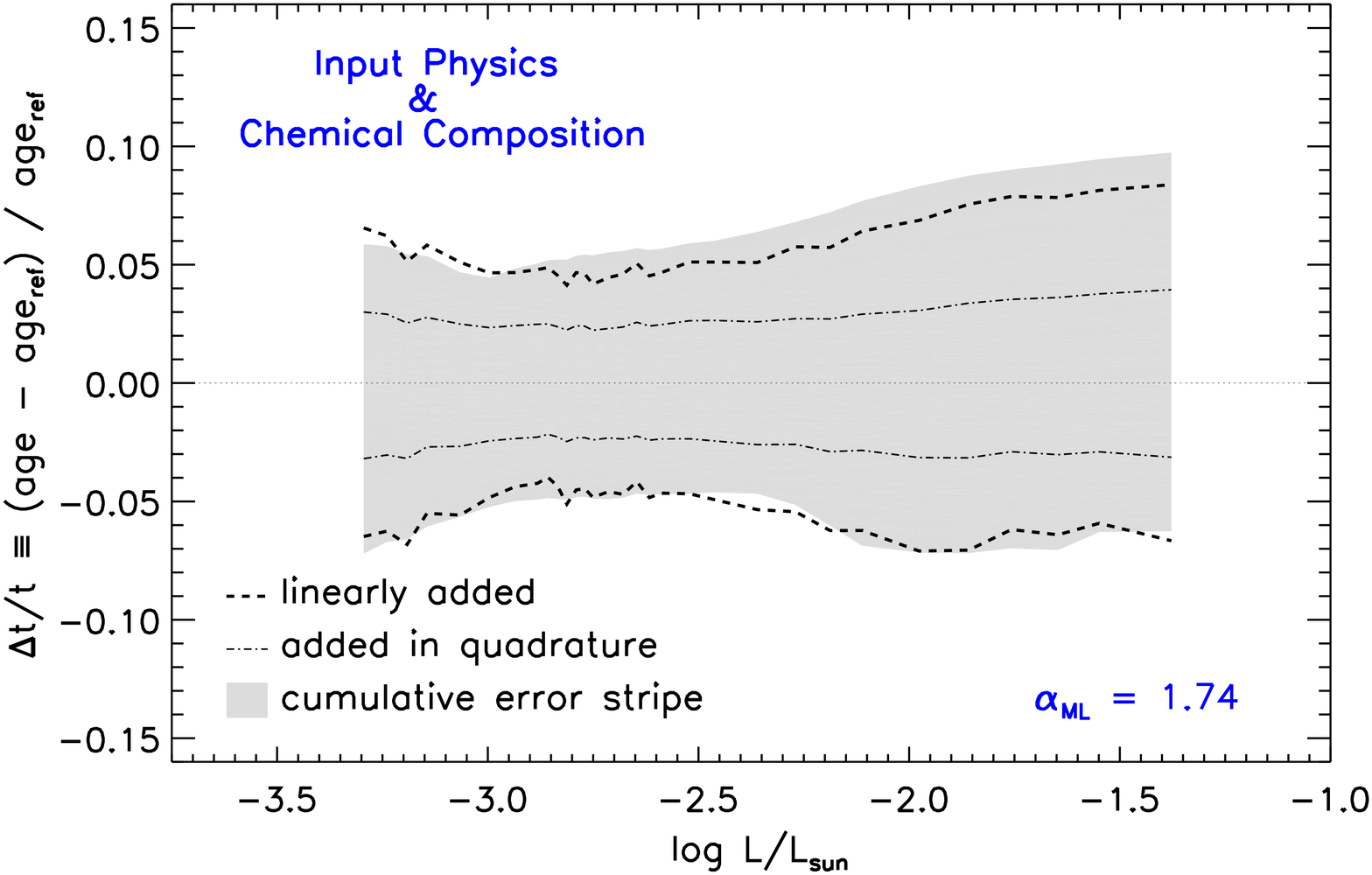}
	\caption{Cumulative error stripe on the relative \ldb{} age computed by taking into account the stellar models with simultaneously perturbed input physics and chemical composition, for \ml = 1.00 (upper panel) and \ml = 1.74 (bottom panel). Errors computed by linearly and quadratically adding uncertainties due to the independent variation of each input physics/chemical composition are overplotted (thick dashed and thin dot-dashed line, respectively).}
	\label{fig:tot_chm_fis}
\end{figure}
Our main aim is to quantify the cumulative uncertainty resulting from input physics and chemical composition, keeping fixed \ml. However, the \ml{} value is itself very uncertain and it might significantly affect the estimate of the cumulative uncertainty. For this reason, Fig. \ref{fig:tot_chm_fis} shows the error stripe on the relative \ldb{} age computed by taking into account the stellar models with simultaneously perturbed input physics and chemical composition for two fixed \ml{} values, namely \ml = 1.00 and 1.74. 

The cumulative error stripe is asymmetric and it gets progressively broader an broader at increasing luminosity. The case with  \ml{}=1.00 shows the largest and most asymmetric stripe. The minimum relative \ldb{} age uncertainty is of about $\pm$5\% while the maximum is of $+$15\%/$-$8\%. The case with \ml{}=1.74 (solar calibrated) shows a thinner and more symmetric error stripe, with a maximum extension of about $+$10\%/$-$8\%.

Both the stripe asymmetry and dependence on \ml{} result essentially from the variation of $\tau_\mathrm{ph}$, which assumes non symmetric values with respect to the reference one (see Sect. \ref{sec:tau}). Indeed, we verified that if the contribution of $\tau_\mathrm{ph}$ is removed from the cumulative uncertainty estimate, the error stripe becomes essentially symmetric and unaffected by \ml. 

Fig. \ref{fig:tot_chm_fis} also shows the relative \ldb{} age uncertainty computed by linearly/quadratically adding the errors due to the individual variation of each input physics/chemical composition used to build the stripe. To do this we computed additional sets of models to account for the effect of the individual variation of each parameter used to obtain $Y$ and $Z$ (i.e. [Fe/H], $\Delta Y / \Delta Z$, and $(Z/X)_\odot$), as listed in Table \ref{tab:chimica_usata_indi}. Such models have been obtained by keeping fixed the input physics and varying a single chemical input within its uncertainty range (i.e. [Fe/H] $\pm0.1$, $\Delta Y / \Delta Z \pm 1$, and $(Z/X)_\odot \pm 15\%$) in equations  (\ref{eq:elio}) and (\ref{eq:metal}). The comparison is intended to quantify the difference between the classical method (linear/quadratic individual error sum), often used in the literature, and the simultaneous variation on a fixed grid (cumulative error stripe) presented in this paper. It is evident that the adoption of a linear sum produces an uncertainty region much closer to the cumulative error stripe than that obtained by a quadratic sum. In particular the latter method gives a systematic underestimation of the uncertainty region (about two times smaller). Fig. \ref{fig:tot_chm_fis} has been computed perturbing simultaneously the input physics and the chemical composition. However, to better understand the results it is worth to discuss the two effects separately.

First, we verified that the contribution of the input physics (keeping the chemical composition fixed to the reference value) sums linearly, as already obtained by \citet{valle13a}. In other words, in the case of a constant chemical composition and small perturbations of the input physics, the uncertainty region obtained by a linear sum coincides with the edges of the cumulative error stripe, while the quadratic sum results in a significative underestimate. 

Secondly, to obtain the uncertainty region edges due to the perturbed chemical composition using both the linear and the quadratic sum, we used the additional models listed in Table \ref{tab:chimica_usata_indi}. It is important to notice that equations (\ref{eq:elio}) and (\ref{eq:metal}) are non-linear in the parameters [Fe/H], $\Delta Y / \Delta Z$ and $(Z/X)_\odot$. In addition, even if a symmetric perturbation of the parameters is adopted the resulting $Y$ and $Z$ variation is not necessarily symmetric. The non-linearity of eqs (\ref{eq:elio}) and (\ref{eq:metal}) introduces a non linear response of the \ldb{} age variation that essentially depends on the initial helium abundance and metallicity perturbation ($\delta Y$, $\delta Z$) with respect to the reference values. In the case of the simultaneous perturbation of all the chemical parameters the initial helium abundance and metallicity variations are simply given by $\delta Y_\rmn{cu.} = Y_\mathrm{ref} - Y_\mathrm{cumulative}$ and $\delta Z_\rmn{cu.} = Z_\mathrm{ref} - Z_\mathrm{cumulative}$ ($Y_\mathrm{cumulative}$ and $Z_\mathrm{cumulative}$ are the values listed in Table \ref{tab:chimica_usata}), while in the case of the individual parameter perturbation by $\delta Y_\rmn{in.} = \sum_{\delta p_j}[Y_\mathrm{ref} -  Y_\mathrm{individual}(\delta p_j)]$ and $\delta Z_\rmn{in.} = \sum_{\delta p_j}[Z_\mathrm{ref} - Z_\mathrm{individual}(\delta p_j)]$ ($Y_\mathrm{individual}(\delta p_j)$ and $Z_\mathrm{individual}(\delta p_j)$ are the values listed in Table \ref{tab:chimica_usata_indi} and $\delta p_j$ is the $j\mathrm{th}$-parameter variation). The non-linearity of equations (\ref{eq:elio}) and (\ref{eq:metal}) leads to $(\delta Y_\rmn{cu.}$, $\delta Z_\rmn{cu.}) \neq (\delta Y_\rmn{in.}$, $\delta Z_\rmn{in.})$, thus to a different variation of the \ldb{} age if the cumulative or independent linear sum models are considered. We verified that this produces the differences between the linear sum edges and the cumulative stripe. 
\begin{figure}
	\centering
	\includegraphics[width=\columnwidth]{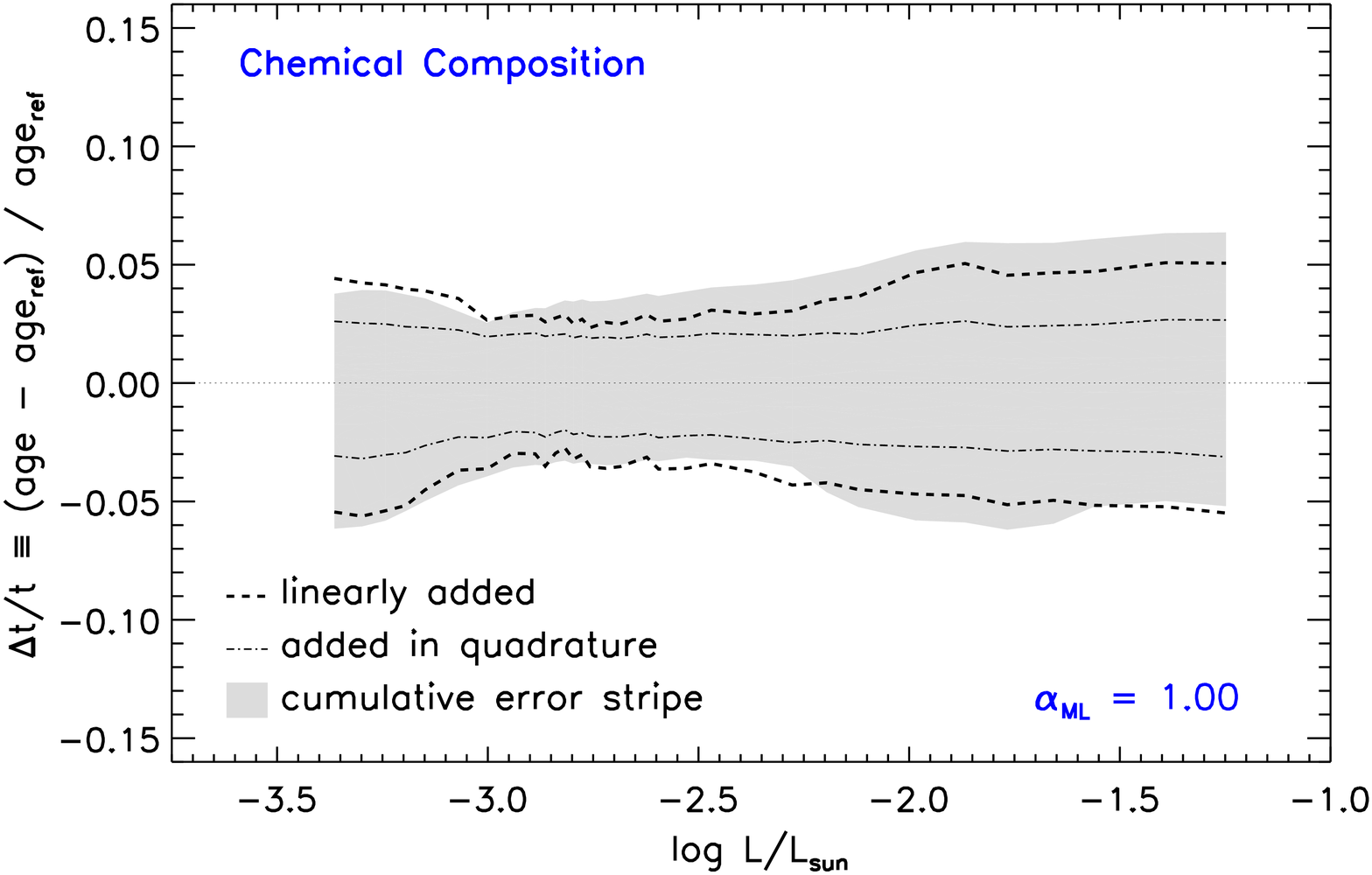}
	\includegraphics[width=\columnwidth]{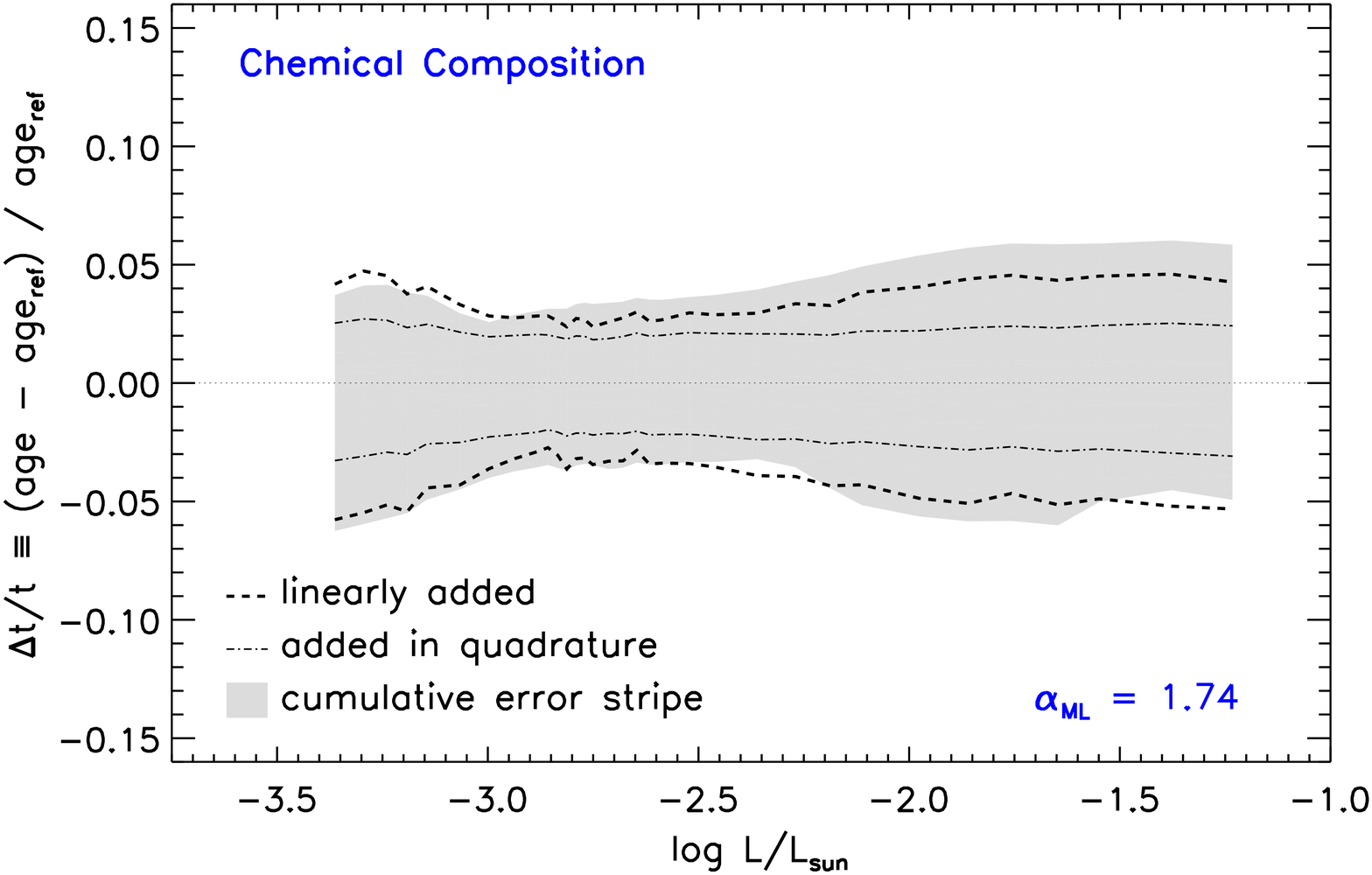}
	\caption{As in Fig. \ref{fig:tot_chm_fis} but taking into account only chemical composition uncertainties.}
	\label{fig:tot_chm}
\end{figure}

Fig. \ref{fig:tot_chm} shows the cumulative error stripe obtained by varying the chemical composition parameters but keeping fixed the input physics to the reference ones, for $\alpha_\mathrm{ML} = 1.00$ and $1.74$. The uncertainty in the initial chemical composition produces a not negligible variation of the \ldb{} age, which is slightly asymmetric due to the asymmetric variation of $Y$ and $Z$, regardless the adopted mixing length. The relative \ldb{} age uncertainty ranges from about 3\% to about 6\%, increasing at larger luminosities. When compared to the total error stripe (which includes also the input physics perturbation), the contribution of the sole uncertainty on the initial chemical composition is of the order of $40\%$. Fig. \ref{fig:tot_chm} also shows the linear/quadratic sum edges. As anticipated, the uncertainty obtained with a linear sum does not coincides with the error stripe even if the differences are quite small (about 1\%), while the quadratic sum drastically underestimates the uncertainty (up to 1.5$-$2 times smaller).

One should be aware that the global error stripe shown in Fig. \ref{fig:tot_chm_fis} is probably an underestimate of the actual uncertainty in the \ldb{} age since, as explained in the previous sections, some potential contributions have been neglected, given the difficulty to properly account for them. However, assuming a linear behaviour also for these contributions, we can give a rough estimate to the total \ldb{} age uncertainty taking into account the additional effect of the outer \bcs, the \eos{}, and the radiative opacity. As shown in Sect. \ref{sec:var_indi_fis}, an uncertainty of 1\% is due to the adoption of a different non-grey \bc{} (Fig. \ref{fig:BCmod}), $1-2\%$ to the \eos{} (Fig. \ref{fig:eos}), and 1\% to an opacity variation (Fig. \ref{fig:ks66_kappa}). Thus, adding these contributions the uncertainty gets larger, ranging from a minimum of about $\pm$9\% to a maximum of $+$19\%/$-$12\%, for $\alpha_\mathrm{ML}=1.00$, or to a maximum of $+$14\%/$-$12\%, for $\alpha_\mathrm{ML} =1.74$. It would be worth to have an estimate of the uncertainty affecting the plasma electron screening factor, which, as shown in Sect. \ref{sec:sk}, might significantly affect the \ldb{} age. Moreover, as already mentioned, we focused on standard \pms{} models that do not take into account some potentially important physical mechanism. In particular it has been recently shown that magnetic fields and star spots might significantly affect the \ldb{} age estimate \citep{jeffries14,malo14}. The systematic bias due to neglecting these physical processes might be comparable or even larger than the cumulative error shown in Fig. \ref{fig:tot_chm_fis}.

\begin{figure}
	\centering
	\includegraphics[width=\columnwidth]{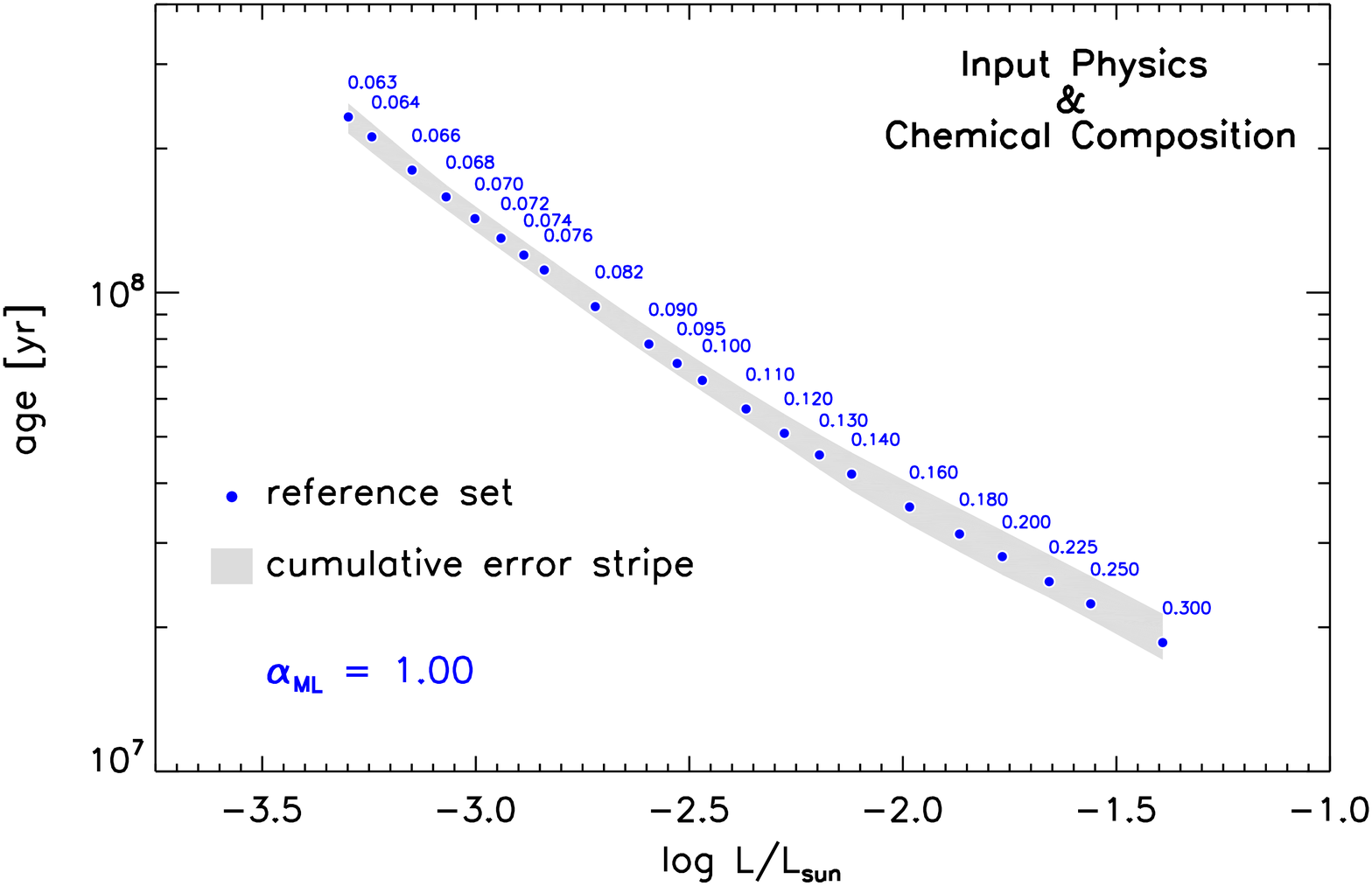}
	\includegraphics[width=\columnwidth]{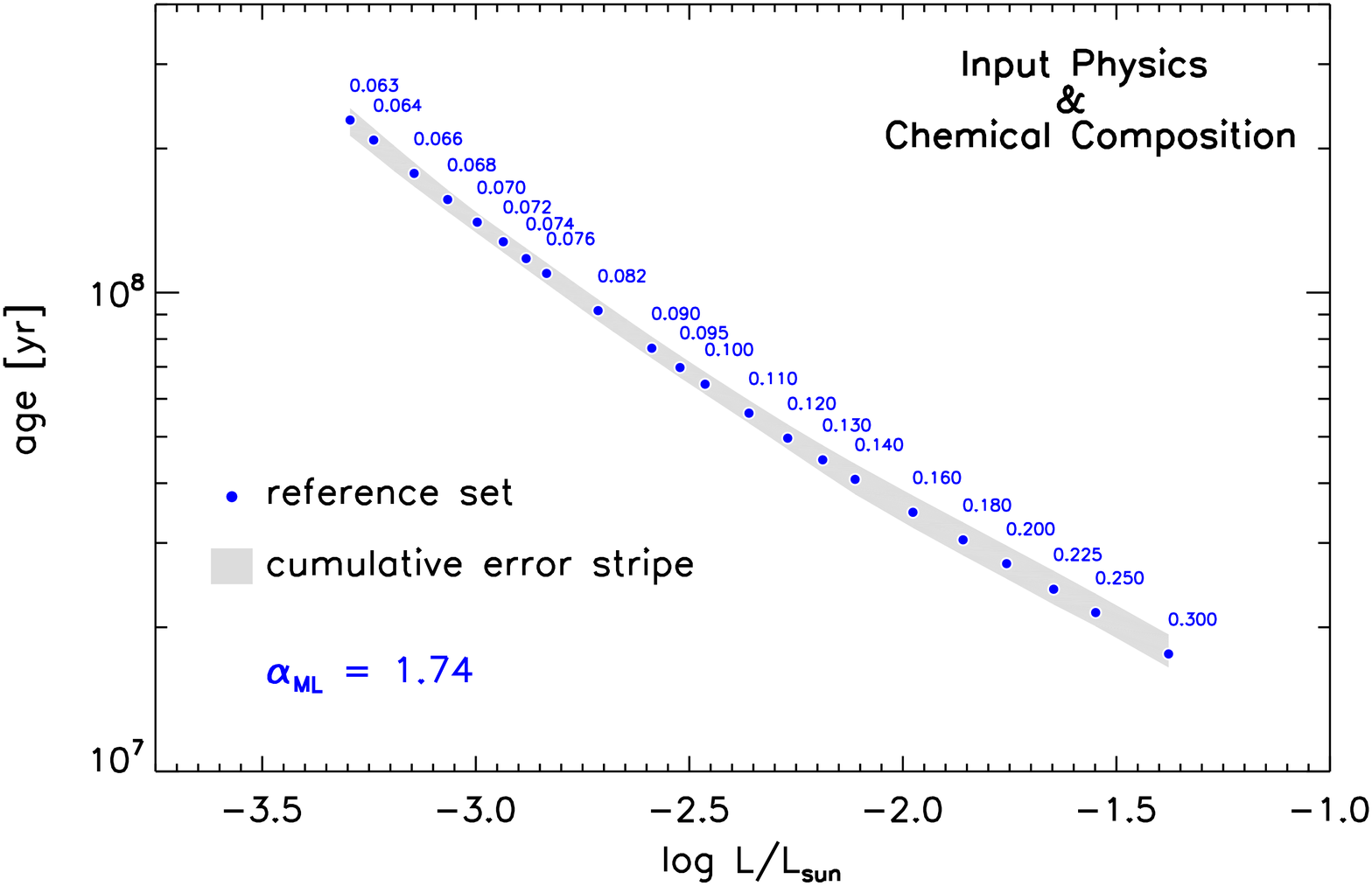}
	\caption{Age at the \ldb{} as a function of luminosity for our reference set of models, with $Z=0.0130$, $Y=0.274$, $X_\mathrm{d}=2\times10^{-5}$, for \ml = 1.00 (upper panel) and \ml=1.74 (bottom panel). The cumulative error stripe is overplotted as a shaded area.}
	\label{fig:tot_logL}
\end{figure}
Finally, Fig. \ref{fig:tot_logL} shows our reference \ldb{} curve in the ($\log L/$L$_\odot$, age) plane with overplotted the cumulative error stripe, for the \ml=1.00 (\emph{upper panel}) and \ml=1.74 set of models (\emph{bottom panel}).

As a final comment, our cumulative \ldb{} age uncertainty is larger (about $50\%$) than that provided by \citet{burke04}. However, a detailed comparison of the two results is not straightforward because of both the different uncertainty sources taken into account and the adopted estimate method. As an example of the former point, we accounted for the uncertainty in the adopted chemical composition, which is lacking in \citet{burke04}. Concerning the latter point, they adopted as best estimate of the total uncertainty a quadratic summation rather than the cumulative stripe described here.

\section{Conclusions}
\label{sec:conclusions}
In this work, we performed a systematic analysis of the main theoretical uncertainties affecting the age at the \ldb{}. With respect to other studies in the literature \citep[e.g.][]{bildsten97,ushomirsky98,jeffries01b,burke04}, which analysed the effect of changing an input physics at the time by keeping fixed all the others, the main novelty of our approach consists in taking into account the simultaneous variation of the main input/parameters without an a priori assumption of independence. More in detail, we computed sets of \pms{} models for all the possible combinations of the perturbations of the ingredients allowed to vary. Such an approach is much more robust in presence of interactions between the varied quantities, but it clearly requires the computation of a huge number of models. For this work, we computed about 12\,000 \pms{} evolutionary tracks.

Besides the method, we improved the analysis of the \ldb{} age uncertainty by studying physical error sources never discussed before, such as the plasma electron screening and the nuclear cross-sections ($^7$Li+p $^2$H+$^2$H, $^2$H+p). A further improvement with respect to previous studies is the detailed computation of the uncertainties propagation due to the adopted initial chemical elements abundances, i.e. the total metallicity, the helium and deuterium abundances, and heavy elements mixture.

The results of the uncertainty computations are shown in Fig. \ref{fig:tot_logL} where the \ldb{} age$-$luminosity curve with the cumulative error stripe for both \ml = 1.00 and 1.74 has been plotted. To the best of our knowledge, this is the first time that such an error stripe has been shown. 

As discussed in \citet{burke04}, the \ldb{} age estimate is more accurate at faint luminosity. Our detailed computations confirm that the error stripe gets progressively broader at increasing luminosities. Moreover, we also showed that the cumulative error is asymmetric and it depends on the adopted \ml{} value. More in detail, the set of models with \ml{}=1.00 shows a larger and more asymmetric error stripe, with positive relative age errors being about twice the negative ones, ranging from $\approx 5$ to $\approx 15\%$. The uncertainty  range reduces to a maximum of about $10\%$ if the solar-calibrated mixing length value (\ml{} = 1.74) is adopted. A further contribution of the order of $4\%$ should be added to the global error to take into account the uncertainty in the \eos{}, outer \bc{} and radiative opacity, not explicitly accounted for in the error stripe computation. Finally, an additional uncertainty source potentially sizeable and worth to be studied in more detail is the electron screening factor. We proved that increasing it of $50\%$ ($100\%$) leads to a $3\%$ ($5.5\%$) variation in the \ldb{} age.

We showed that the cumulative error stripe computed by simultaneously perturbing the input physics is in excellent agreement with the simple linear sum of the \ldb{} relative age differences obtained by individually perturbing the input physics themselves. Regarding the chemical composition uncertainties, the agreement is worse but always better than $1\%$. On the contrary, the use of a quadratic sum (as generally done in the literature) results in a systematic underestimate of the total \ldb{} age uncertainty, which is about 1.5/2 times smaller than that obtained with the cumulative error stripe.

Another result of this work consists in quantifying, for the first time, the effect of the initial chemical elements abundances uncertainty on the \ldb{} age. We showed that the contribution of the sole chemical composition uncertainty is not negligible ranging from $\pm$ 3 to $\pm$ 6\%, thus accounting for at least $\approx 40\%$ of the total error budget; this part of the error stripe is almost symmetric and independent of \ml{} value. 

\section*{Acknowledgements}
We would like to thank the anonymous referee for the valuable comments that helped us to improve the paper. This work has been supported by PRIN-MIUR 2010-2011 (\emph{Chemical and dynamical evolution of the Milky Way and Local Group galaxies}, PI F. Matteucci) and  PRIN-INAF 2012 (\emph{The M4 Core Project with Hubble Space Telescope}, PI L. Bedin). 
%
%

\bibliographystyle{mn2e}
\bibliography{bibliography}
\label{lastpage}
\end{document}